\documentclass[aps,prd,preprint,groupedaddress, nofootinbib]{revtex4-1}
\usepackage{bm}
\usepackage{mathrsfs}
\usepackage{color}
\usepackage{tikz}
\usepackage{enumerate}
\usepackage{amsthm}
\usepackage{pgfplots}
\usepackage{graphicx}
\usepackage{epstopdf}

\theoremstyle{definition}

\begin{document}


\title{Gedanken Experiments to Destroy a Black Hole II: Kerr-Newman Black Holes Cannot be Over-Charged or Over-Spun}


\author{Jonathan Sorce}
\affiliation{Department of Physics, University of Chicago, Chicago, Illinois 60637, USA}
\author{Robert M. \surname{Wald}}
\affiliation{Enrico Fermi Institute and Department of Physics, University of Chicago, Chicago, Illinois 60637, USA}


\date{\today}

\begin{abstract}
We consider gedanken experiments to destroy an extremal or nearly extremal Kerr-Newman black hole by causing it to absorb matter with sufficient charge and/or angular momentum as compared with energy that it cannot remain a black hole. It was previously shown by one of us that such gedanken experiments cannot succeed for test particle matter entering an extremal Kerr-Newman black hole. We generalize this result here to arbitrary matter entering an extremal Kerr-Newman black hole, provided only that the non-electromagnetic contribution to the stress-energy tensor of the matter satisfies the null energy condition. We then analyze the gedanken experiments proposed by Hubeny and others to over-charge and/or over-spin an initially slightly non-extremal Kerr-Newman black hole. Analysis of such gedanken experiments requires that we calculate all effects on the final mass of the black hole that are second-order in the charge and angular momentum carried into the black hole, including all self-force effects. We obtain a general formula for the full second order correction to mass, $\delta^2 M$, which allows us to prove that no gedanken experiments of the generalized Hubeny type can ever succeed in over-charging and/or over-spinning a Kerr-Newman black hole, provided only that the non-electromagnetic stress-energy tensor satisfies the null energy condition. Our analysis is based upon Lagrangian methods, and our formula for the second-order correction to mass is obtained by generalizing the canonical energy analysis of Hollands and Wald to the Einstein-Maxwell case. Remarkably, we obtain our formula for $\delta^2 M$ without having to explicitly compute self-force or finite size effects. Indeed, in an appendix, we show explicitly that our formula incorporates both the self-force and finite size effects for the special case of a charged body slowly lowered into an uncharged black hole.
\end{abstract}

\pacs{}

\maketitle

\section{Introduction}
\label{sec:intro}
The Kerr-Newman family of metrics are the unique stationary, asymptotically flat black hole solutions of the Einstein-Maxwell equations in 4 spacetime dimensions. The Kerr-Newman metrics comprise a 3-parameter family of solutions parameterized by mass $M$, charge $Q$, and angular momentum $J = M a$. However, these solutions describe black holes only for a limited region of this parameter space, characterized by the inequality
\begin{equation}
	M^2
		\geq (J/M)^2 + Q^2. \label{CC_inequality}
\end{equation}
When this inequality is not satisfied, the spacetime contains a naked singularity, i.e., the singularity is visible from infinity.

The above facts give rise to a possible means of testing the weak cosmic censorship conjecture \cite{penrose1969}, \cite{wald1997}, which states that all singularities arising from gravitational collapse must be hidden within black holes, so that no physical process can give rise to a naked singularity. Suppose that we start with a Kerr-Newman black hole satisfying (\ref{CC_inequality}). Now throw/drop matter into the black hole carrying energy $E$, angular momentum, $\ell$, and charge $q$, so that the final state will have mass $M+E$, angular momentum $J+\ell$, and charge $Q+q$. Then if $\ell$ and/or $q$ can be made sufficiently large compared with $E$, the inequality (\ref{CC_inequality}) will be violated, resulting in a contradiction with the final state being a black hole.

The most obvious case to consider for an attempt to destroy a black hole in this manner would be to start with an extremal black hole, satisfying $M^2 = (J/M)^2 + Q^2$, and to throw in particle matter. This case was analyzed in 1974 by one of us in paper I of this series \cite{wald1974}. It was shown in paper I that no violations of (\ref{CC_inequality}) can occur by throwing particle matter into an extremal Kerr-Newman black hole. The nature of this result is well illustrated by considering the special case of attempting to ``over-charge'' an extremal Reissner-Nordstrom ($Q=M$) black hole. Let $\xi^a$ denote the horizon Killing field, which, for a Reissner-Nordstrom black hole, coincides with the static Killing field $(\partial / \partial t)^a$. A test particle with mass $m$ and charge $q$ in this spacetime has energy given by
\begin{equation}
	E
		= - (m u_a + q A_a) \xi^a \, ,
\label{energy}
\end{equation}
where $u_a$ is the four-velocity of the particle and $A_a$ is the vector potential of the black hole's electromagnetic field.

Since $\xi^a$ is null on the horizon, the first term $-m u_a \xi^a$ is non-negative on the horizon, although it can be made arbitrarily small. Thus, the energy of a particle that crosses the horizon is bounded below by the electromagnetic potential energy term
\begin{equation}
	E
		\geq q \Phi_H, 
		\label{RN_particle_bound}
\end{equation}
where $\Phi_H = (- A_a \xi^a)|_H$ is the electromagnetic potential evaluated on the horizon. However, $\Phi_H = 1$ for an extremal Reissner-Nordstrom black hole, so any particle that enters the black hole must satisfy
\begin{equation}
	E
		\geq q \, .
\label{RN1}
\end{equation}
Consequently, we have $M + E \geq Q + q$, so (\ref{CC_inequality}) holds. In other words, any particle with sufficiently large charge $q$ as compared with $E$ to produce a violation of (\ref{CC_inequality}) for the final state would be repelled by the electric field of the black hole and thus cannot enter it. As shown in paper I \cite{wald1974}, similar results hold for attempting to over-charge and/or over-spin a general extremal Kerr-Newman black hole using particle matter.

Nevertheless, in 1999 Hubeny \cite{hubeny1999} proposed that violations of (\ref{CC_inequality}) might still occur if one suitably added matter to a slightly non-extremal black hole. To see this, consider a slightly non-extremal Reissner-Nordstrom black hole. It is useful to introduce the dimensionless parameter
\begin{equation}
	\epsilon
		= \frac{\sqrt{M^2 - Q^2}}{M} \, ,
\label{eps}
\end{equation}
so that $\epsilon \rightarrow 0$ in the extremal limit. For $\epsilon \ll 1$, we have 
\begin{equation}
	\Phi_H = Q / r_+ \approx 1 - \epsilon,
\end{equation}
where $r_+ = M + \sqrt{M^2 - Q^2}$ is the horizon radius. In place of (\ref{RN1}) we now obtain
\begin{equation}
	E
		\geq q (1 - \epsilon) \, .
\label{RN2}
\end{equation}
Consequently, for this lower bound for $E$, we have
\begin{equation}
(M+E) - (Q+q) \approx - \epsilon q + \frac{M \epsilon^2}{2}  \, .
\label{RN_ccv}
\end{equation}
Thus, it might appear that we can obtain a violation of (\ref{CC_inequality}) by taking $q > \epsilon M / 2$ (but still keeping $q \ll Q$). 

The main difficulty with Hubeny's argument is that for $q \sim \epsilon M$, the violation of (\ref{CC_inequality}) given by (\ref{RN_ccv}) is of order $\epsilon q \sim q^2/M$. Consequently, to determine if one truly can obtain a violation of (\ref{CC_inequality}), the quantities appearing in (\ref{RN_ccv}) must all be calculated consistently to the appropriate order. Specifically, the energy, $E$, of the matter must be calculated to order $q^2$. However, formula (\ref{energy}) applies only to ``test matter'' and is valid only to linear order in $q$; it does not take into account the contributions of electromagnetic self-energy (which require consideration of bodies of finite size) or the energy contributed by self-force effects, both of which enter at order $q^2$. In particular, it is possible that self-force effects could contribute to a repulsion of the body from the black hole, requiring that the body be given additional energy at order $q^2$ in order to enter the black hole. 

Similar potential violations of (\ref{CC_inequality}) have been found for Reissner-Nordstr\"{o}m black holes absorbing angular momentum \cite{felice2001}, Kerr black holes absorbing charge or angular momentum \cite{hod2002,jacobson2009,sotiriou2010}, and for generic Kerr-Newman black holes \cite{saa2011, gao2013}. However, just as in Hubeny's argument, in order to determine whether these potential violations actually occur, one needs to calculate all contributions to energy that are quadratic order in the relevant parameters of the particle. This would appear to require a complete analysis of self-force effects as well as finite size effects and any other effects that might enter at this order.

Unfortunately, the analytic computation of electromagnetic and gravitational self-force effects on the motion of bodies near a Kerr-Newman black hole is well beyond present capabilities. Thus, the main results that have been obtained thus far have come from numerical simulations. Numerical work has indicated that the self-force on particles falling into black holes may suffice to prevent Hubeny-type violations from occurring in the specific cases of over-charging a nearly extremal Reissner-Nordstr\"{o}m black hole \cite{poisson2013} and over-spinning a nearly extremal Kerr black hole \cite{barausse2010, barausse2011, barack2015A, barack2015B}. However, even for these special cases, no general analysis has been given of the second order corrections to energy. As such, there is no general proof that the cosmic censorship inequality (\ref{CC_inequality}) holds at quadratic order for processes involving matter that falls into nearly extremal Kerr-Newman black holes.

The main purpose of this paper is to give a complete analysis---valid to second order---of the contributions to the mass of a black hole for arbitrary matter that enters a black hole. At linear order, we derive a general expression---first obtained in \cite{wald2001}---that expresses $\delta M$ in terms of the flux of charge and angular momentum carried into the black hole together with the non-electromagnetic energy flux. Assuming only that the non-electromagnetic contribution to the stress energy tensor satisfies the null energy condition, we will prove that for arbitrary processes involving matter falling into an exactly extremal Kerr-Newman black hole, no violation of (\ref{CC_inequality}) can occur at linear order in the perturbation. This result, which was previously obtained for charged scalar matter in \cite{toth2012} and generalized in \cite{natario2016}, generalizes the results derived for particle matter in paper I \cite{wald1974} to completely general matter. 

We then consider the possible Hubeny-type violations that might occur for slightly non-extremal black holes. Our general formula for $\delta M$ shows that the linear order process obeys a generalization of (\ref{RN2}), thus allowing the possibility of a violation of (\ref{CC_inequality}) but requiring an analysis of the second order effects on energy. We will perform this analysis by expressing the second order change in mass, $\delta^2 M$, of the black hole in terms of the canonical energy of the first order perturbation. We will then make the additional assumption that the \textit{non-extremal} black hole is stable under \textit{linear} perturbations, so that the first order perturbation decays to a stationary final state. This will allow us to evaluate the canonical energy in terms of a positive flux contribution through the horizon and a contribution from the final stationary perturbation. The resulting formula gives rise to an inequality on $\delta^2 M$, and we will see that this inequality is just what is needed to prove that no violations of the Hubeny type can ever occur. Remarkably, we are able to derive this inequality---which automatically takes account of all self-force and finite size effects---without having to explicitly calculate these effects themselves. We will show by explicit calculation in the Appendix that for the special case of lowering a charged body into an uncharged black hole, our general formula corresponds precisely to taking these effects into account.

Our analysis differs from most previous analyses---including that of paper I \cite{wald1974}---in the following three key respects: (1) We consider completely general matter rather than particle matter. Of course, ``particle matter'' makes sense in general relativity only when considered to be a limiting case of general matter as described in \cite{gralla_wald_1} and \cite{gralla_wald_2}, so the general results derived in this paper also automatically hold for physically realizable particle matter. (2) Rather than analyzing the motion of bodies to determine what trajectories will or will not enter the black hole, we simply restrict consideration to the case where all matter that is initially present enters the black hole, and we compute the second order variation of the mass for this case. This allows us to derive the desired inequality without having to calculate the motion of bodies. (3) Most importantly, we obtain an exact expression for the full second order effects on the mass of a black hole. This allows us to obtain the above-mentioned inequality on $\delta^2 M$.

In section \ref{sec:variational_identities}, we obtain the general variational formulas that we will need, including the generalization of the notion of canonical energy introduced in \cite{wald2012} for vacuum perturbations of vacuum black holes to the Einstein-Maxwell case. The gedanken experiments to destroy an extremal black hole are analyzed in section \ref{sec:extremal}. We consider a perturbation of the black hole involving matter with charge and angular momentum such that the black hole is initially unperturbed in a neighborhood of the horizon and such that all of the matter eventually falls into the black hole. We obtain a general expression for $\delta M$ that was first derived in \cite{wald2001}. We show that this expression yields an inequality that is sufficient to show that no violations can occur at linear order for extremal black holes, as previously found in \cite{natario2016}. This generalizes the results of paper I to completely general matter whose non-electromagnetic stress-energy satisfies the null energy condition. The Hubeny-type gedanken experiments to destroy a slightly non-extremal black hole are considered in section \ref{sec:nonextremal}. We consider a process that is optimal at first order so that the first order perturbation saturates our lower bound on $\delta M$. We obtain an expression for $\delta^2 M$ involving the canonical energy of the first order perturbation. Assuming that the first order perturbation of the non-extremal black hole becomes stationary at late times (i.e., that the non-extremal black hole is linearly stable), we obtain a lower bound on $\delta^2 M$ that is sufficient to prove that no violations of (\ref{CC_inequality}) can occur. A simple pictorial representation of our results is presented in section \ref{sec:discussion}. The relationship between our results and the electromagnetic self-force and self-energy is detailed in the Appendix for the case of a charged body lowered into an uncharged black hole.

Our metric signature, curvature, and abstract index conventions follow \cite{wald_book}. In many instances, we will suppress the indices on differential forms, in which case they will be denoted with boldface letters.

\section{Variational Identities and Canonical Energy for Einstein-Maxwell Theory}
\label{sec:variational_identities}

In this section, we generalize the canonical energy results obtained in \cite{wald2012} for vacuum perturbations of vacuum black holes to the Einstein-Maxwell case. It would be most natural to treat the electromagnetic field $A_a$ as a connection on a principal $U(1)$-bundle and use the framework developed by Prabhu \cite{prabhu2017} for doing the Lagrangian analysis in the principal bundle. However, since this would require the introduction of considerable machinery and formalism, we will bypass this here and simply treat $A_a$ as the one-form that one obtains on spacetime by making a choice of gauge. This leads to some awkwardness in that we will work---as is conventional---in a gauge such that, in the background black hole spacetime, $A_a$ is stationary, $\pounds_\xi A_a = 0$, and $A_a \rightarrow 0$ at infinity, so the ``horizon potential'' $\Phi_H = - \xi^a A_a |_{\mathscr{H}}$ is non-vanishing, where $\xi^a$ is the horizon Killing field and $\mathscr{H}$ denotes the future event horizon. Since $\xi^a = 0$ on the bifurcation surface, this implies that, in our gauge, $A_a$ cannot be smooth at the bifurcation surface as a one-form on spacetime, which might be thought to cause difficulties. In fact, no such difficulties occur, as can be seen by performing the analysis in the principal bundle in the framework of Prabhu \cite{prabhu2017}. Namely, the connection, $A_a$, is smooth as a one-form in the bundle and this is consistent with the non-vanishing of $\Phi_H$ because the lift of $\xi^a$ to the bundle has non-vanishing vertical part. Nevertheless, to keep our discussion simple, we will perform our analysis on spacetime and ignore the non-smoothness of the background $A_a$, relying on the fact that the analysis could have been performed in the principal bundle, where all fields are smooth.

Although our interest is in $4$-dimensional Kerr-Newman black holes in Einstein-Maxwell theory, we will consider general diffeomorphism covariant theories in $n$-dimensional spacetimes in subsections \ref{subsec:linear} and \ref{subsec:canonical_energy}. In \ref{subsec:linear}, we review the derivation of a fundamental variational identity for theories derived from a diffeomorphism covariant Lagrangian. We define canonical energy in \ref{subsec:canonical_energy}. The Einstein-Maxwell case in $4$ spacetime dimensions is explicitly considered in \ref{subsec:EM_theory}. Gauge invariance issues are treated in \ref{subsec:gauge}.

\subsection{The Linear Variational Identity}
\label{subsec:linear}

The Lagrangian for a diffeomorphism-covariant theory on an $n$-dimensional spacetime is given by an $n$-form $\mathbf{L}$ on spacetime, which is a local function of the metric, $g_{ab}$, its curvature, and symmetrized covariant derivatives of the curvature, and which may also depend on other tensor fields, $\psi$, and their symmetrized covariant derivatives. We refer to the full field configuration as $\phi = (g_{ab}, \psi)$. We vary the Lagrangian by considering a one-parameter family of field configurations, $\phi(\lambda)$, and taking derivatives of $\mathbf{L}$ with respect to $\lambda.$ Throughout this paper, the notation ``$\delta$'' will be used to denote derivatives evaluated at $\lambda = 0$, e.g.,
\begin{equation}
	\delta \mathbf{L}
		= \left. \frac{d \mathbf{L}}{d\lambda} \right|_{\lambda = 0},
	\quad
	\delta^2 \mathbf{L}
		= \left. \frac{d^2 \mathbf{L}}{d\lambda^2} \right|_{\lambda = 0},
	\quad
	\delta \phi
		= \left. \frac{d \phi}{d\lambda} \right|_{\lambda = 0}.
\end{equation}

The first-order variation of the Lagrangian can be written as
\begin{equation}
	\frac{d \mathbf{L}}{d\lambda}
		= \mathbf{E}(\phi) \cdot \frac{d \phi}{d\lambda} + d \bm{\theta}\left( \phi, \frac{d \phi}{d\lambda} \right), \label{lagrangian_first_variation}
\end{equation}
where $\mathbf{E}$ is locally constructed from the fields $\phi$ and their derivatives, while $\bm{\theta}$ is locally constructed from $\phi, d \phi/d\lambda$, and their derivatives; $\bm{\theta}$ corresponds to the ``boundary term'' one would obtain by putting the variation of $\mathbf{L}$ under an integral sign and integrating by parts to remove all spacetime derivatives from $d \phi/d \lambda$. The Euler-Lagrange equations of motion of the theory are simply
\begin{equation}
	\mathbf{E}(\phi)
		= 0 \, .
\end{equation}

The \emph{symplectic current} $(n-1)$-form $\bm{\omega}$ is defined in terms of a second variation of $\bm{\theta}.$ For a two-parameter family of field configurations $\phi(\lambda_1, \lambda_2)$, we define
\begin{equation}
	\bm{\omega}\left(\phi; \frac{\partial \phi}{\partial \lambda_1}, \frac{\partial \phi}{\partial \lambda_2} \right)
		= \frac{\partial}{\partial \lambda_1} \bm{\theta}\left(\phi, \frac{\partial \phi}{\partial \lambda_2} \right)
			- \frac{\partial}{\partial \lambda_2} \bm{\theta}\left(\phi, \frac{\partial \phi}{\partial \lambda_1} \right).
\end{equation}
The symplectic current depends on the background field configuration $\phi$, as well as on the perturbations $\partial \phi/\partial \lambda_1$ and $\partial \phi/\partial \lambda_2$. If both of these perturbations satisfy the linearized equations of motion $\frac{\partial}{\partial \lambda_1} E(\phi) = \frac{\partial}{\partial \lambda_2} E(\phi) = 0$, then it follows from equation (\ref{lagrangian_first_variation}) that
\begin{equation}
	d \bm{\omega} = 0, \label{symplectic_current_conservation}
\end{equation}
i.e., the symplectic current is conserved.

The Noether current associated with an arbitrary vector field $X^a$ is defined as
\begin{equation}
	\bm{\mathcal{J}}_{X}(\phi)
		= \bm{\theta}(\phi; \mathscr{L}_{X} \phi) - \iota_X \mathbf{L}(\phi) \, ,
\label{nc}
\end{equation}
where $\iota_X \mathbf{L}$ denotes contraction of $X^a$ into the first index of the differential form $\mathbf{L}.$ A simple calculation \cite{wald1994} shows that the first variation of $\bm{\mathcal{J}}_X$ can be written as
\begin{eqnarray}
	\frac{d \bm{\mathcal{J}}_X}{d\lambda}
		&=& - \iota_X\left(\mathbf{E}(\phi) \cdot \frac{d \phi}{d\lambda} \right)
			+ \bm{\omega}\left(\phi; \frac{d \phi}{d\lambda}, \mathscr{L}_X \phi \right) \nonumber \\*
			&& + d\left[\iota_X \bm{\theta}\left(\phi, \frac{d \phi}{d\lambda} \right) \right]. \label{varied_Noether_current}
\end{eqnarray}
On the other hand, it was shown in \cite{wald1995} that the Noether current can be written in the form
\begin{equation}
	\bm{\mathcal{J}}_X
		= \mathbf{C}_X + d \mathbf{Q}_X, \label{Noether_current_charge_form}
\end{equation}
where $\mathbf{Q}_X$ is called the \emph{Noether charge} and $\mathbf{C}_X \equiv X^a \mathbf{C}_a$ are the constraints of the theory, so that $\mathbf{C}_a = 0$ when the equations of motion are satisfied. In particular, $d \bm{\mathcal{J}} = 0$ when the equations of motion are satisfied, as can be shown directly from the definition (\ref{nc}) of $\bm{\mathcal{J}}$.

By differentiating\footnote{Note that we take $X^a$ to be $\lambda$-independent.} equation (\ref{Noether_current_charge_form}) with respect to $\lambda$ and comparing it to equation (\ref{varied_Noether_current}), we obtain the fundamental identity
\begin{eqnarray}
	d \left[ \frac{d \mathbf{Q}_X}{d\lambda} - \iota_X \bm{\theta} \left(\phi, \frac{d \phi}{d\lambda} \right) \right]
		&=& \bm{\omega}\left(\phi; \frac{d \phi}{d\lambda}, \mathscr{L}_X \phi \right) - \frac{d \mathbf{C}_X}{d\lambda} \nonumber \\*
			&&
			- \iota_X\left(\mathbf{E}(\phi) \cdot \frac{d \phi}{d\lambda} \right) \, .
\label{fundamental_linear_identity}
\end{eqnarray}
This identity forms the basis for all calculations conducted in the remainder of this paper. 

Now, assume that $\phi(\lambda)$ is globally hyperbolic with Cauchy surface $\Sigma$. Evaluating  (\ref{fundamental_linear_identity}) at $\lambda = 0$ and integrating the resulting equation over $\Sigma$, we obtain
\begin{eqnarray}
	\int_{\partial \Sigma} \left[ \delta \mathbf{Q}_X - \iota_X \bm{\theta} \left(\phi, \delta \phi \right) \right]
		&=& \int_{\Sigma} \bm{\omega}\left(\phi; \delta \phi, \mathscr{L}_X \phi \right) -  \int_{\Sigma} \delta \mathbf{C}_X \nonumber \\*
			&& -  \int_{\Sigma} \iota_X\left(\mathbf{E}(\phi) \cdot \delta \phi \right) \, .
\label{intid}
\end{eqnarray}
A Hamiltonian $h_X$ associated with a vector field $X^a$ is a functional of $\phi$ such that if and only if $\phi$ satisfies the equations of motion, then under all variations $\delta \phi$ we have
\begin{equation}
	\delta h_{X}
		= \int_\Sigma \bm{\omega}\left(\phi; \delta \phi, \mathscr{L}_X \phi \right).
\end{equation}
If the spacetime is asymptotically flat and there is no ``interior boundary'' to $\Sigma$, then a Hamiltonian, $h_X$, conjugate to $X^a$ must satisfy
\begin{equation}
	\delta h_{X}
		=  \int_{\infty} \left[ \delta \mathbf{Q}_X - \iota_X \bm{\theta} \left(\phi, \delta \phi \right) \right]
		+ \int_{\Sigma} \delta \mathbf{C}_X \, ,
\end{equation}
where ``$\int_{\infty}$'' denotes the limit to spatial infinity of integration over a suitable family of spacelike $(n-2)$-spheres.
This motivates the following definition\footnote{We assume here that the matter fields fall off at infinity rapidly enough so as not to contribute to the surface integral on the right side of (\ref{adm}). Otherwise, these matter fields may make contributions of the form ``potential times varied charge'' that would need to be subtracted to obtain the conventional definition of ADM conserved quantities.} of the ADM conserved quantity $H_X$ conjugate to an asymptotic symmetry $X^a$ for asymptotically flat solutions: $H_X$ (if it exists) is the quantity such that, for all one-parameter families of solutions, we have
\begin{equation}
	\delta H_{X}
		= \int_{\infty} \left[ \delta \mathbf{Q}_X - \iota_X \bm{\theta} \left(\phi, \delta \phi \right) \right] \, .
\label{adm}
\end{equation}
 
Finally, let us restrict consideration to the case where (i) $\phi_0 = \phi(\lambda = 0)$ is a globally hyperbolic, asymptotically flat solution of the equations of motion, $\mathbf{E} = 0$, and (ii) $\phi_0$ possesses a Killing field $\xi^a$ that is also a symmetry of the matter fields $\psi$, so that $\mathscr{L}_{\xi} \phi_0 = 0$. Then (\ref{intid}) yields
\begin{equation}
	\int_{\partial \Sigma} \left[ \delta \mathbf{Q}_\xi - \iota_\xi \bm{\theta} \left(\phi, \delta \phi \right) \right]
		=  -  \int_{\Sigma} \delta \mathbf{C}_\xi \, .
\label{intid2}
\end{equation}
The case of greatest interest for us is where $\phi_0$ represents the exterior of a stationary black hole, and $\xi^a$ is the horizon Killing field
\begin{equation}
\xi^a = t^a + \Omega_H \varphi^a \, ,
\end{equation}
where $t^a$ is the timelike Killing field of $\phi_0$, $\varphi^a$ is the axial Killing field of $\phi_0$, and $\Omega_H$ is the angular velocity of the horizon. The contribution to the boundary integral from infinity is then just
\begin{equation}
\int_{\infty} \left[ \delta \mathbf{Q}_\xi - \iota_\xi \bm{\theta} \left(\phi, \delta \phi \right) \right] = \delta H_\xi = \delta M - \Omega_H \delta J,
\label{bndyinf}
\end{equation}
where $M$ is the ADM mass and $J$ is the ADM angular momentum.
If the spacetime represents the exterior of a black hole, then there will be a contribution from the ``internal boundary'' as well. We will evaluate this internal boundary contribution for Einstein-Maxwell theory in subsection C below.

\subsection{Second Order Variations and Canonical Energy}
\label{subsec:canonical_energy}

Let us now continue to restrict consideration to the case where $\phi_0 = \phi(\lambda = 0)$ is a globally hyperbolic solution of the equations of motion that possesses a Killing field $\xi^a$ that is also a symmetry of the matter fields $\psi$, so that $\mathscr{L}_{\xi} \phi_0 = 0$. Again, we do {\em not} require that the perturbation $\delta \phi = (d \phi/d \lambda)|_{\lambda = 0}$ satisfy the linearized equations of motion. Let $\Sigma$ be a Cauchy surface. We define the \emph{canonical energy} of the perturbation $\delta \phi$ on $\Sigma$ by
\begin{equation}
	\mathcal{E}_{\Sigma}(\phi; \delta \phi)
		\equiv \int_{\Sigma} \bm{\omega}\left(\phi; \delta \phi, \mathscr{L}_{\xi} \delta \phi \right).
\label{canen}
\end{equation}

We can obtain an extremely useful expression for canonical energy by differentiating (\ref{fundamental_linear_identity}) with respect to $\lambda$ and evaluating the resulting expression at $\lambda = 0$. We obtain
\begin{eqnarray}
	d \left[ \delta^2 \mathbf{Q}_{\xi} - \iota_{\xi} \delta \bm{\theta} \left(\phi, \delta \phi \right) \right]
		&=& \bm{\omega}\left(\phi; \delta \phi, \mathscr{L}_{\xi} \delta \phi \right) - \delta^2 \mathbf{C}_{\xi} \nonumber \\*
			&& - \iota_{\xi} \left(\delta \mathbf{E} \cdot \delta \phi \right),
		\label{canonical_energy_density}
\end{eqnarray}
Here, the meaning of the ``$\delta$'s'' in the expression $\delta \bm{\theta}(\phi, \delta \phi)$ is that both derivatives in this term are to be evaluated simultaneously, i.e.,
\begin{equation}
	\delta \bm{\theta}(\phi, \delta \phi)
		\equiv \left. \left[\frac{d}{d\lambda} \bm{\theta}\left(\phi, \frac{d\phi}{d \lambda}\right)\right]\right|_{\lambda=0}.
\end{equation}
Integrating (\ref{canonical_energy_density}) over $\Sigma$, we obtain
\begin{eqnarray}
	\mathcal{E}_{\Sigma}(\phi; \delta \phi)
		& = & 	\int_{\partial \Sigma} \left[ \delta^2 \mathbf{Q}_{\xi} - \iota_{\xi} \delta \bm{\theta} \left(\phi, \delta \phi \right) \right]
				+ \int_{\Sigma} \delta^2 \mathbf{C}_{\xi} \nonumber \\*
			&& 	+ \int_{\Sigma} \iota_{\xi} \left(\delta \mathbf{E} \cdot \delta \phi \right).
	\label{canonical_energy_identity}
\end{eqnarray}

The case we are most interested in here is one where $\phi_0$ corresponds to a stationary black hole, $\xi^a$ is the horizon Killing field,\footnote{Note that in \cite{wald2012}, the canonical energy was defined with respect to the asymptotically timelike Killing field $t^a$ rather than the horizon Killing field $\xi^a$. These quantities are equal to each other for axisymmetric perturbations, as considered in \cite{wald2012}.} 
and $\Sigma$ is a Cauchy surface for the exterior of the black hole. In that case, it follows from (\ref{adm}) that the contribution to the 
the boundary term in (\ref{canonical_energy_identity}) from infinity is
\begin{equation}
	\int_{\infty} \left[ \delta^2 \mathbf{Q}_{\xi} - \iota_{\xi} \delta \bm{\theta} \left(\phi, \delta \phi \right) \right] = \delta^2 M - \Omega_H \delta^2 J \, .
\label{d2M}
\end{equation}
We will evaluate the interior boundary term at the end of the next subsection.

\subsection{Einstein-Maxwell Theory}
\label{subsec:EM_theory}

We now consider Einstein-Maxwell theory in $4$ spacetime dimensions and provide explicit expressions for many of the quantities appearing in the previous subsections. The Einstein-Maxwell Lagrangian is given by
\begin{equation}
	\mathbf{L}
		= \frac{1}{16\pi} (R - F^{ab} F_{ab}) \bm{\epsilon},
\end{equation}
where $\bm{\epsilon}$ is the volume element associated with the metric. For this Lagrangian, the field configuration consists of the metric and the vector potential, $\phi = (g_{ab}, A_a)$. As explained in the introductory paragraph to this section, we will treat $A_a$ as a one-form on spacetime. The symplectic potential, Noether charge, equations of motion, and constraints for this Lagrangian were computed in \cite{wald2001}. The symplectic potential can be written as
\begin{equation}
\theta_{abc}\left(\phi, \frac{d \phi}{d\lambda} \right)= \theta^{GR}_{abc} + \theta^{EM}_{abc},
\label{EM_boundary_term}
\end{equation}
where
\begin{eqnarray}
	\theta^{GR}_{abc}\left(\phi, \frac{d \phi}{d\lambda} \right)
		& = & \frac{1}{16\pi} \epsilon_{dabc} g^{de} g^{fg} \nonumber \\*
		&& \times \left( \nabla_g \frac{d g_{ef}}{d \lambda} - \nabla_e \frac{d g_{fg}}{d \lambda} \right) \label{GR_boundary_term} \\
	\theta^{EM}_{abc}\left(\phi, \frac{d \phi}{d\lambda} \right)
		& = & - \frac{1}{4\pi} \epsilon_{dabc} F^{de} \frac{d A_e}{d \lambda}. \label{electromagnetic_boundary_term}
\end{eqnarray}
The Noether charge is given by 
\begin{equation}
(Q_{X})_{ab} = (Q_X^{GR})_{ab} + (Q_{X}^{EM})_{ab},
\label{EM_Noether_charge}
\end{equation}
where
\begin{eqnarray}
	(Q_{X}^{GR})_{ab}
		& = & - \frac{1}{16\pi} \epsilon_{abcd} \nabla^c X^d, \label{GR_Noether_charge} \\
	(Q_X^{EM})_{ab}
		& = & - \frac{1}{8\pi} \epsilon_{abcd} F^{cd} A_e X^e. \label{electromagnetic_Noether_charge} 
\end{eqnarray}
The equations of motion and constraints are given by
\begin{eqnarray}
	\mathbf{E}(\phi) \cdot \frac{d \phi}{d\lambda}
		& = & - \bm{\epsilon} \left[ \frac{1}{2} T^{ab} \frac{d g_{ab}}{d\lambda} + j^a \frac{d A_a}{d\lambda} \right], \label{EM_equations_of_motion} \\
	C_{bcda}
		& = & \epsilon_{ebcd} \left[ T_{a}{}^{e} + A_a j^e \right]. \label{EM_constraints}
\end{eqnarray}
Here we have written $T_{ab} \equiv G_{ab} - 8 \pi T^{EM}_{ab}$---so that $T_{ab}$ corresponds to the non-electromagnetic part of the stress-energy tensor, and $j^a = (1/4\pi) \nabla_b F^{ab}$---so that $j^a$ corresponds to the electromagnetic charge-current. Note that in the absence of sources, when both $T_{ab}$ and $j_a$ are zero, the constraints (\ref{EM_constraints}) vanish and the Euler-Lagrange equations of motion (\ref{EM_equations_of_motion}) are satisfied.

The symplectic current for the Einstein-Maxwell theory can be written in the form
\begin{equation}
	\omega_{abc}\left(\phi; \frac{\partial \phi}{\partial \lambda_1}, \frac{\partial \phi}{\partial \lambda_2} \right) = \omega^{GR}_{abc} + \omega^{EM}_{abc},
\label{EM_symplectic_current}
\end{equation}
where, from equation (\ref{EM_boundary_term}), we have
\begin{eqnarray}
	\omega^{GR}_{abc}
		& = & \frac{1}{16\pi} \epsilon_{dabc} w^d, \label{GR_symplectic_current}  \\
	\omega^{EM}_{abc}
		& = & \frac{1}{4 \pi} \left[ \frac{\partial}{\partial \lambda_2} (\epsilon_{dabc} F^{de}) \frac{\partial A_e}{\partial \lambda_1} \right. \nonumber \\*
		&& \left.
							- \frac{\partial}{\partial \lambda_1} (\epsilon_{dabc} F^{de}) \frac{\partial A_e}{\partial \lambda_2} \right],
							\label{electromagnetic_symplectic_current}
\end{eqnarray}
where, in (\ref{GR_symplectic_current}), we have
\begin{equation}
	w^a
		= P^{abcdef}
			\left( \frac{\partial g_{bc}}{\partial \lambda_2} \nabla_d \frac{\partial g_{ef}}{\partial \lambda_1}
				- \frac{\partial g_{bc}}{\partial \lambda_1} \nabla_d \frac{\partial g_{ef}}{\partial \lambda_2} \right),
\end{equation}
with
\begin{eqnarray}
	P^{abcdef}
		&=& g^{ae} g^{fb} g^{cd} - \frac{1}{2} g^{ad} g^{be} g^{fc} - \frac{1}{2} g^{ab} g^{cd} g^{ef} \nonumber \\*
		&& 	- \frac{1}{2} g^{bc} g^{ae} g^{fd} + \frac{1}{2} g^{bc} g^{ad} g^{ef}.
\end{eqnarray}

We now restrict attention to the case where  $\phi_0 = \phi(\lambda = 0)$ is a stationary black hole solution to the Einstein-Maxwell equations (i.e., $T^{ab} = j^a = 0$ at $\lambda = 0$) with horizon Killing field $\xi^a$, and we let $\Sigma$ be a Cauchy surface for the exterior region. In fact, by the black hole uniqueness theorems \cite{wald_book}, $\phi_0$ must be a Kerr-Newman solution, but we need not make use of this fact here. We work in a gauge where $\mathscr{L}_{\xi} A_a(\lambda = 0) = 0$ and $A_a(\lambda=0) \rightarrow 0$ at infinity. As already discussed in the first paragraph of this section, in this gauge, $A_a(\lambda=0)$ will, in general, be singular at the horizon, but this does not cause any difficulties. Furthermore, the variations $\delta A_a$ and $\delta^2 A_a$ may be assumed to be smooth (as can be justified by working in the principal bundle framework of Prabhu \cite{prabhu2017}).

By definition, for a {\em non-extremal} black hole the horizon will be of bifurcate type, and $\Sigma$ will terminate at the bifurcation surface $B$. For a non-extremal black hole, we now evaluate the boundary contribution to (\ref{intid2}) arising from $B$. Since $\xi^a = 0$ on $B$, we have
\begin{equation}
\int_{B} \left[ \delta \mathbf{Q}^{GR}_{\xi} - \iota_{\xi} \bm{\theta}^{GR}(\phi, \delta \phi) \right] = 
\int_{B} \delta \mathbf{Q}^{GR}_{\xi} 
= \frac{\kappa}{8 \pi} \delta A_B,
\end{equation}
where $A_B$ is the area of $B$ and $\kappa$ is the surface gravity of the event horizon.
To evaluate the electromagnetic contribution to the boundary term\footnote{We assume that $A_a t^a$ and $A_a \varphi^a$ fall off as $1/r$ and $F_{ab}$ falls off as $1/r^2$ at infinity, so there is no electromagnetic contribution to the boundary term at infinity.} at $B$, we note that by (\ref{electromagnetic_boundary_term}), $\bm{\theta}^{EM}$ is smooth at $B$ (since $\delta A_a$ is smooth), so $\iota_{\xi} \bm{\theta}^{EM} = 0$. However, by (\ref{electromagnetic_Noether_charge}), we have
\begin{equation}
	\delta \mathbf{Q}^{EM}_{\xi}
		=  - \frac{1}{8\pi} \left[ \xi^e A_e \delta (\epsilon_{abcd} F^{cd}) +  \xi^e (\delta A_e) \epsilon_{abcd} F^{cd}) \right].
\end{equation}
Again, the second term vanishes at $B$ on account of the smoothness of $\delta A_a$ and the vanishing of $\xi^a$. However, the quantity
\begin{equation}
\Phi_H \equiv - \left[\xi^e A_e(\lambda) \right] |_{\mathscr{H}}
\end{equation}
is, in general, nonvanishing at $B$. Since $\Phi_H$ must be constant on the horizon at $\lambda=0$ \cite{carter_book} (see theorem 1 of \cite{prabhu2017} for a general proof for Yang-Mills fields), we find that the electromagnetic contribution to the boundary term at $B$ is 
\begin{eqnarray}
	\int_{B} \left[ \delta \mathbf{Q}^{EM}_{\xi} - \iota_{\xi} \bm{\theta}^{EM}(\phi, \delta \phi) \right]
		& = &  \frac{1}{8\pi} \Phi_H \int_B \delta (\epsilon_{abcd} F^{cd}) \nonumber \\*
		& = & \Phi_H \delta Q_B,
\end{eqnarray}
where $Q_B$ is the electric charge flux integral over $B$. 

The ingredients are now in place to write out (\ref{intid2}) explicitly for a non-extremal black hole. We previously evaluated the boundary term from infinity in (\ref{bndyinf}), and, in the previous paragraph, we have evaluated the boundary term from $B$. Using (\ref{EM_constraints}) and the fact that $T_{ab} = j^a = 0$ in the background spacetime (since $\phi_0$ is a solution), we see that the remaining term $\delta \mathbf{C}_\xi$ takes the form
\begin{equation}
\delta C_{bcda} \xi^a
		 =  \epsilon_{ebcd} \left[ \delta T_{a}{}^{e} + A_{a} \delta j^e \right] \xi^a
\end{equation}
Thus, we see that (\ref{intid2}) takes the explicit form
\begin{widetext}
\begin{equation}
\delta M - \Omega_H \delta J - \frac{\kappa}{8 \pi} \delta A_B - \Phi_H \delta Q_B 
		 =  - \int_\Sigma \epsilon_{ebcd} \left[ \delta T_{a}{}^{e} + A_{a} \delta j^e \right] \xi^a.
\label{intid3}
\end{equation}
\end{widetext}
For source free perturbations, $\delta T_{ab} = \delta j_a = 0$, this yields the usual first law of black hole mechanics of Einstein-Maxwell theory.

It should be emphasized that (\ref{intid3}) holds only for non-extremal black holes. In this paper, we will be concerned with both non-extremal and extremal black holes. However, it is clear from the derivation that (\ref{intid3}) (with $\delta A_B = \delta Q_B = 0$) also holds for extremal black holes in the special case where $\Sigma$ is not a Cauchy surface but rather an asymptotically flat hypersurface with one boundary at spatial infinity and the other boundary on the horizon at an early time such that the perturbation vanishes in a neighborhood of this internal boundary. In this case, there clearly will be no boundary contribution from the internal boundary of $\Sigma$. We will use (\ref{intid3}) in this form for extremal black holes in section \ref{sec:extremal}.

The canonical energy may also be split into gravitational and electromagnetic contributions
\begin{equation}
\mathcal{E}_{\Sigma}(\phi; \delta \phi) = \mathcal{E}_{\Sigma}^{GR} + \mathcal{E}_{\Sigma}^{EM} \, .
\end{equation}
Explicit formulas for these parts can be obtained from the definition (\ref{canen}), substituting from (\ref{GR_symplectic_current}) and (\ref{electromagnetic_symplectic_current}). These formulas are quite complicated and will not be written out explicitly here. Fortunately, we will need to evaluate the canonical energy integral only over (a portion of) the horizon (where its form simplifies considerably) and for stationary perturbations (where it can be evaluated straightforwardly). 

We may now explicitly evaluate the terms appearing in (\ref{canonical_energy_identity}) for Einstein-Maxwell theory, in exact parallel with our above evaluation of the terms appearing in (\ref{intid2}). For a non-extremal black hole, we obtain\footnote{It should be noted that since we take $\xi^a$ to be fixed, the quantities $\Omega_H$ and $\kappa$ do not vary. This means that if we perturb toward another stationary black with different values of $\Omega_H$ or $\kappa$, then $\xi^a$ cannot be the horizon Killing field of the perturbed black hole. See \cite{wald2012} for further discussion.}
\begin{widetext}
\begin{equation}
	\delta^2 M - \Omega_H \delta^2 J - \Phi_H \delta^2 Q_B - \frac{\kappa}{8 \pi} \delta^2 A_B
		= \mathcal{E}_{\Sigma}(\phi; \delta \phi) - \int_{\Sigma} \iota_{\xi} (\delta \mathbf{E}(\phi) \cdot \delta \phi) - \int_{\Sigma} \delta^2 \mathbf{C}_{\xi}.
		\label{EM_canonical_energy_identity}
\end{equation}
\end{widetext}
Again, this equation (with $\delta^2 A_B = \delta^2 Q_B = 0$) will hold for an extremal black hole if we restrict consideration to the case where both the first and second order perturbations vanish in a neighborhood of the horizon at the internal boundary of $\Sigma$. In section \ref{sec:nonextremal}, we will evaluate the right side of (\ref{EM_canonical_energy_identity}) in the context relevant to our calculations.

\subsection{Gauge Invariance of Canonical Energy}
\label{subsec:gauge}

In this subsection, we show that the canonical energy is gauge invariant when evaluated on linearized solutions to the Einstein-Maxwell equations, subject to the restrictions of Proposition 1 below. It should be noted that the symplectic form (i.e., the integral of $\bm{\omega}(\phi, \delta_1 \phi, \delta_2 \phi)$ over a Cauchy surface) is \textit{not} gauge invariant, either in the sense of the Maxwell gauge transformations $\delta A_a \mapsto \delta A_a + \nabla_a \chi$ or the infinitesimal diffeomorphisms $\delta \phi \mapsto \delta \phi + \mathscr{L}_X \phi$, on account of boundary terms arising from the horizon.

For the purposes of analyzing gauge invariance, it is convenient to view the canonical energy as a bilinear form on the space of perturbations to a black hole background given by
\begin{equation}
	\mathcal{E}_{\Sigma}(\phi; \delta_1 \phi, \delta_2 \phi)
		\equiv \int_{\Sigma} \bm{\omega}(\phi; \delta_1 \phi, \mathscr{L}_{\xi} \delta_2 \phi).
		\label{Equad}
\end{equation}
The canonical energy will be gauge invariant if and only if it vanishes whenever $\delta_1 \phi$ or $\delta_2 \phi$ is a pure gauge transformation.

If $\delta_1 \phi$ and $\delta_2 \phi$ are solutions, then, as shown in \cite{wald2012}, $\mathcal{E}_{\Sigma}$ is symmetric. Namely,
by the antisymmetry and bilinearity of the symplectic current, we have
\begin{equation}
    \mathcal{E}_{\Sigma}(\phi; \delta_1 \phi, \delta_2 \phi) - \mathcal{E}_{\Sigma}(\phi; \delta_2 \phi, \delta_1 \phi)
        = \int_{\Sigma} \mathscr{L}_{\xi} \bm{\omega}(\phi; \delta_1 \phi, \delta_2 \phi).
\end{equation}
Applying the Lie derivative identity $\mathscr{L}_{\xi} \bm{\omega} = \iota_{\xi} d \bm{\omega} + d (\iota_{\xi} \bm{\omega})$ and applying Stokes' theorem to the second term yields
\begin{widetext}
\begin{equation}
    \mathcal{E}_{\Sigma}(\phi; \delta_1 \phi, \delta_2 \phi) - \mathcal{E}_{\Sigma}(\phi; \delta_2 \phi, \delta_1 \phi)
        = \int_{\Sigma} \iota_{\xi} d \bm{\omega}(\phi; \delta_1 \phi, \delta_2 \phi)
            + \int_{\infty} \iota_{\xi} \bm{\omega}(\phi; \delta_1 \phi, \delta_2 \phi)
            - \int_{B} \iota_{\xi} \bm{\omega}(\phi; \delta_1 \phi, \delta_2 \phi).
\end{equation}
\end{widetext}
The first term vanishes for solutions\footnote{The perturbations considered in sections \ref{sec:extremal} and \ref{sec:nonextremal} do \textit{not} satisfy the linearized equations of motion, since they have sources in the form of charged matter that is added to the black hole. However, the quantity $\int_{\Sigma} \iota_{\xi} d \bm{\omega}$ still vanishes for the particular surface $\Sigma$ chosen in those sections (cf. figures \ref{fig:extremal_matter} and \ref{fig:nonextremal_matter}), and so the gauge invariance established in this subsection still holds for that particular case.} by (\ref{symplectic_current_conservation}). The boundary term at infinity vanishes under the assumption that $\delta_1 \phi$ and $\delta_2 \phi$ are asymptotically flat perturbations with appropriate falloff conditions and the boundary term at the bifurcation surface vanishes since $\xi^a$ vanishes on $B$, thus establishing that $\mathcal{E}_{\Sigma}$ is symmetric. This is convenient because it implies that to show gauge invariance of $\mathcal{E}_{\Sigma}$, we need only show that $\mathcal{E}_{\Sigma}$ vanishes when $\delta_2 \phi$ is pure gauge in (\ref{Equad}).

First let us consider a pure Maxwell gauge transformation given by $\delta g_{ab} = 0, \delta A_a = \nabla_a \chi$ for some smooth function $\chi$. In analogy with (\ref{nc}), which defined the Noether current associated with a local diffeomorphism, we may define the Noether current associated with a Maxwell gauge transformation by
\begin{equation}
    \bm{\mathscr{J}}_{\chi}
        = \bm{\theta}(\phi, \nabla_a \chi).
\end{equation}
Just as in (\ref{Noether_current_charge_form}), this Noether current can also be written in terms of a constraint and a charge as
\begin{equation}
    \bm{\mathscr{J}}_\chi
        = \bm{\mathcal{C}}[\chi] + d \bm{\mathcal{Q}}[\chi].
\end{equation}
A simple calculation shows that for the Einstein-Maxwell theory, the constraint and Noether charge are given by
\begin{eqnarray}
    (\mathcal{C}[\chi])_{abc}
        & = & \epsilon_{dabc} \chi j^d, \label{Maxwell_gauge_constraints} \\
    (\mathcal{Q}[\chi])_{ab}
        & = & - \frac{1}{8 \pi} \epsilon_{cdab} \chi F^{cd}. \label{Maxwell_gauge_charge}
\end{eqnarray}

A calculation similar to that used to obtain (\ref{intid}) yields the identity
\begin{equation}
    \int_{\partial \Sigma} \delta \bm{\mathcal{Q}}[\chi]
        = \int_{\Sigma} \bm{\omega}(\phi; \delta \phi, \nabla_a \chi)
            - \int_{\Sigma} \delta \bm{\mathcal{C}},
\end{equation}
i.e.,
\begin{equation}
    W_{\Sigma}(\phi; \delta \phi, \nabla_a \chi)
        = \int_{\infty} \delta \bm{\mathcal{Q}}[\chi] - \int_B \delta \bm{\mathcal{Q}}[\chi] + \int_{\Sigma} \delta \bm{\mathcal{C}},
\end{equation}
where $W_{\Sigma}(\phi; \delta_1 \phi, \delta_2 \phi) \equiv \int_{\Sigma} \bm{\omega}(\phi; \delta_1 \phi, \delta_2\phi)$ is the symplectic form. The constraint term vanishes under the assumption that $\delta \phi$ satisfies the linearized equations of motion, so, using (\ref{Maxwell_gauge_charge}), we obtain,
\begin{eqnarray}
	W_{\Sigma}(\phi; \delta \phi, \nabla_a \chi)
		& = & -\frac{1}{8 \pi} \int_\infty \chi \delta(\epsilon_{cdab} F^{cd}) \nonumber \\*
		&& + \frac{1}{8 \pi} \int_B \chi \delta(\epsilon_{cdab} F^{cd}). \label{Maxwell_gauge_sym_form}
\end{eqnarray}
This expression is nonvanishing for generic perturbations and gauge transformations, since $\chi$ may be non vanishing at infinity and at $B$. Thus, the symplectic form is not invariant under Maxwell gauge transformations. However, the gauge invariance of the canonical energy for Maxwell gauge transformation can be seen by replacing $\chi$ by $\mathscr{L}_{\xi} \chi = \xi^a \nabla_a \chi$ in (\ref{Maxwell_gauge_sym_form}). The resulting expression vanishes, since $\xi^a \nabla_a \chi$ goes to zero at infinity and vanishes at $B$.
Thus, the Einstein-Maxwell canonical energy is indeed invariant under Maxwell gauge transformations, as we desired to show.

W now analyze the gauge dependence of the canonical energy under smooth infinitesimal diffeomorphisms, $\delta \phi = \mathscr{L}_{X} \phi$, for which $X^a$ is an asymptotic symmetry.
The canonical energy of an infinitesimal diffeomorphism is given by
\begin{eqnarray}
    \mathcal{E}_{\Sigma}(\phi; \delta \phi, \mathscr{L}_{X} \phi)
        & = & W_{\Sigma}(\phi; \delta \phi, \mathscr{L}_{\xi} \mathscr{L}_{X} \phi) \nonumber \\*
        & = & W_{\Sigma}(\phi; \delta \phi, \mathscr{L}_{Y} \phi),
\end{eqnarray}
where $Y^a = [\xi, X]^a$ and we have used the fact that $\mathscr{L}_{\xi} \phi = 0$ at $\lambda=0.$
From (\ref{intid}) and (\ref{adm}), we have
\begin{eqnarray}
	\mathcal{E}_{\Sigma}(\phi; \delta \phi, \mathscr{L}_{X} \phi)
		& = & W_{\Sigma}(\phi; \delta \phi, \mathscr{L}_{Y} \phi) \nonumber \\*
		& = & \delta H_{Y} - \int_{B} \left[ \delta \mathbf{Q}_Y - \iota_Y \bm{\theta} \left(\phi, \delta \phi \right) \right], \,\,\,\,\,\,
		\label{GR_gauge_sym_form_1}
\end{eqnarray}
where we have used the assumptions that $\phi(\lambda=0)$ and $\delta \phi$ satisfy the equations of motion and the linearized equations of motion, respectively.

It is easily seen that the right side of (\ref{GR_gauge_sym_form_1}) cannot vanish unless some restrictions are placed on the allowed perturbations at the horizon and at infinity. These conditions are purely gauge conditions on the perturbations that do not restrict the physical perturbations we consider. First,
following \cite{wald2012}, we impose the gauge condition that the perturbed expansion of the horizon generators vanishes,
\begin{equation}
\delta \Theta|_{\mathscr{H}} =  0 \, .
\label{expan}
\end{equation}
As shown in \cite{wald2012}, this condition may always be imposed for non-extremal black holes. The infinitesimal diffeomorphisms $X^a$ that preserve this condition are the ones that are tangent to the future horizon. This implies that $Y^a = \mathscr{L}_{\xi} X^a$ is normal to the horizon at $B$.

Second, we impose the condition
\begin{equation}
    k^a \delta A_a|_{\mathscr{H}} =  0 \, ,
\label{Maxwell_gauge_condition}
\end{equation}
where $k^a$ denotes an affinely parametrized tangent to the generators of the horizon.
This condition always can be imposed by a Maxwell gauge transformation $\delta A_a \rightarrow \delta A'_a = \delta A_a - \nabla_a \chi$ with $\chi$ satisfying $k^a \nabla_a \chi = k^a \delta A_a$ on $\mathcal{H}$.

We now evaluate the terms appearing on the right side of (\ref{GR_gauge_sym_form_1}), where $Y^a = \mathscr{L}_{\xi} X^a$. First, we evaluate the contribution to the boundary term at $B$ arising from the symplectic potential.
We split the symplectic potential into a gravitational and an electromagnetic part as in (\ref{EM_boundary_term}). As shown in \cite{wald2012}, the gravitational part of the symplectic potential contribution yields
\begin{equation}
    \int_B \iota_Y \bm{\theta}^{GR}(\phi, \delta \phi)
        = - \frac{1}{8 \pi} \int_B f \delta \Theta \bm{\epsilon},
    \label{GR_gauge_GR_theta}
\end{equation}
where we have written $Y^a = f k^a$ on $B$ with $k^a$ normal to the horizon, since $Y^a$ is normal to the horizon at $B$. This term vanishes as a consequence of our gauge condition (\ref{expan}).

As for the electromagnetic part of the symplectic potential, we have
\begin{equation}
    \int_B \iota_X \bm{\theta}^{EM}(\phi, \delta \phi)
        = - \frac{1}{4 \pi} \int_B \epsilon_{dcab} Y^c F^{de} \delta A_e.
\end{equation}
However, the assumption that the background spacetime is stationary restricts the form of $F^{de}$, since the flux of electromagnetic stress-energy
\begin{equation}
        T_{ab}^{EM}
            = \frac{1}{4\pi} \left[ F_{ac} F_{b}{}^{c} - \frac{1}{4} g_{ab} F^{cd} F_{cd} \right]\label{electromagnetic_stress_energy}
\end{equation}
through the horizon must vanish. For this flux to vanish, we must have $T_{ab}^{EM} k^a k^b = 0$ on the horizon. The dominant energy condition (which is automatically satisfied by the electromagnetic field) then implies that $T_{ab}^{EM} k^a$ must be proportional to $k_b$. This implies that on $\mathscr{H}$,  $F^{ab}$ must take the form
\begin{equation}
    F^{ab} = v^{[a} k^{b]} + w^{ab},
    \label{HEM}
\end{equation}
where $w^{ab}$ is purely tangential to the horizon. From this, and from the assumption that $X^a$ is tangent to the horizon generators on $B$, we find that the electromagnetic part of the symplectic potential can be written as
\begin{equation}
    \int_B \iota_Y \bm{\theta}^{EM}(\phi, \delta \phi)
        = - \frac{1}{8 \pi} \int_B \epsilon_{dcab} Y^c v^d k^e \delta A_e \, ,
    \label{GR_gauge_EM_theta}
\end{equation}
where we have used the fact that the pullback to $\mathscr H$ of $\epsilon_{abcd}$ contracted into any vector tangent to $\mathscr H$ vanishes. The right side of (\ref{GR_gauge_EM_theta}) vanishes on account of our gauge condition (\ref{Maxwell_gauge_condition}).

Next, we consider the term $\delta H_Y$ in (\ref{GR_gauge_sym_form_1}).
Since $X^a$ is an asymptotic symmetry and $\xi^a = t^a + \Omega_H \varphi^a$ for a Kerr-Newman background, $Y^a$ is a linear combination of an asymptotic space translation and an asymptotic rotation or boost in a direction orthogonal to the black hole's axis of rotation. So long as we restrict ourselves to perturbations with vanishing ADM linear momenta, $\delta P_i = 0$, and vanishing ADM angular momentum and center of mass in directions orthogonal to the axis of rotation, we have $\delta H_Y = 0$ for all suitable choices of infinitesimal diffeomorphism $X^a.$ These conditions do not restrict the physical perturbation.

We are left with
\begin{equation}
    \mathcal{E}_{\Sigma}(\phi; \delta \phi, \mathscr{L}_{X} \phi)
        = - \int_{B} \delta \mathbf{Q}_Y.
\end{equation}
We split $\mathbf{Q}_Y$ into gravitational and electromagnetic parts as in (\ref{EM_Noether_charge}).
It was shown in Appendix A of \cite{wald2012} that since $Y^a$ is normal to the horizon, the pullback to $B$ of $\delta \bm{Q}^{GR}_Y$ is given by
\begin{equation}
\delta \bm{Q}^{GR}_Y
= -\frac{1}{16 \pi} (\delta \epsilon_{abcd}) \nabla^c Y^d \, .
\end{equation}
The right side will be nonvanishing if and only if the quantity
\begin{equation}
U \equiv n_{cd} \nabla^c Y^d
\end{equation}
is nonvanishing on $B$ in the background spacetime, where $n_{ab} = n_{[ab]}$ is the binormal to $B$.
We substitute $Y^a = \mathscr{L}_{\xi} X^a = \xi^b \nabla_b X^a - X^b \nabla_b \xi^a$ in this equation and expand using the Leibniz rule to get
\begin{eqnarray}
	U
		& = & n_{cd} \left[\xi^b \nabla^c \nabla_b X^d + (\nabla^c \xi^b) \nabla_b X^d \right. \nonumber \\*
		&& \left. - X^b \nabla^c \nabla_b \xi^d - (\nabla^c X^b) \nabla_b \xi^d   \right].
\end{eqnarray}
The first term vanishes since $\xi^a$ vanishes on $B$. Since $\xi^a$ is a Killing field, we have $\nabla_a \nabla_b \xi^c  = {R^c}_{bad} \xi^d = 0$ on $B$, so the third term also vanishes on $B$. Finally, using the fact that $\nabla_a \xi_b \propto n_{ab}$ on $B$, the second and fourth terms can be seen to cancel. Thus, $U=0$ on $B$ and the contribution from $\delta \bm{Q}^{GR}_Y$ vanishes.

\medskip
\noindent
\textbf{Remark} In \cite{wald2012}, the vanishing of the contribution from $\delta \bm{Q}^{GR}_Y$ was obtained by imposing the gauge condition $\delta \epsilon_{ab} = (\delta A/A) \epsilon_{ab}$ on the area element on $B$ together with the restriction $\delta A = 0$ on the perturbation. The above calculation shows that it was not necessary to impose either this gauge condition or this restriction. In particular, the hypothesis that $\delta A = 0$ may be dropped from Proposition 3 of \cite{wald2012}.
\medskip

Finally, we evaluate the contribution from $\delta \bm{Q}^{EM}_Y$. We obtain
\begin{equation}   
    \int_B \delta \bm{Q}^{EM}_Y    =  - \int_B \frac{1}{8 \pi} \delta(\epsilon_{abcd} F^{cd}) A_e Y^e.
    \label{GR_gauge_CE_penultimate}
\end{equation}
However, a diffeomorphism $X^a$ will preserve our gauge condition (\ref{Maxwell_gauge_condition}) only if $\xi^a \mathscr{L}_{X} A_a = 0$ on the horizon\footnote{Rather than restricting $X^a$ so as to preserve the gauge condition (\ref{Maxwell_gauge_condition}), it would be more sensible to require that any $X^a$ that violates  (\ref{Maxwell_gauge_condition}) be accompanied by a Maxwell gauge transformation that restores  (\ref{Maxwell_gauge_condition}). One would then get a nonvanishing contribution from (\ref{GR_gauge_CE_penultimate}) that would then be canceled by the contribution from the Maxwell gauge transformation.}, which implies that $A_a Y^a$ vanishes at $B$. Thus, the contribution from $\delta \bm{Q}^{EM}_Y$ also vanishes.

We summarize the results of this subsection in the following proposition:

\noindent
\textbf{Proposition 1}:
Consider the subspace of perturbations, $\delta \phi$, that (i) satisfy the linearized equations of motion, $\delta E(\phi) = 0$, (ii) satisfy the gauge conditions (\ref{expan}) and (\ref{Maxwell_gauge_condition}) at the horizon, and (iii) have vanishing ADM linear momenta, $\delta P_i = 0$, and vanishing ADM angular momentum and center of mass in directions orthogonal to the axis of rotation of the unperturbed black hole. Then the Einstein-Maxwell canonical energy $\mathcal{E}_{\Sigma}(\phi; \delta_1 \phi, \delta_2 \phi)$ on this subspace is invariant under all infinitesimal diffeomorphisms $\delta \phi = \mathscr{L}_{X} \phi$ and Maxwell gauge transformations $\delta A_a = \nabla_a \chi$ (where it is understood that these transformations must preserve conditions (ii) and (iii)).

\section{Gedanken Experiments to Destroy an Extremal Black Hole}
\label{sec:extremal}

Consider an extremal Kerr-Newman black hole, 
\begin{equation}
	M^2
		= (J/M)^2 + Q^2 \, .
\end{equation}
We wish to see if we can cause the inequality (\ref{CC_inequality}) to be violated by throwing/dropping charged and/or rotating matter into the black hole. Specifically, (\ref{CC_inequality}) will be violated---and a contradiction with cosmic censorship obtained---if we can perturb the black hole so that
\begin{equation}
	2 M \delta M
		<  2(J/M) (M \delta J - J \delta M)/M^2 + 2 Q \delta Q\, .
\end{equation}
Writing $a = J/M$, we see that a violation will occur if the perturbation satisfies
\begin{equation}
	\delta M
		< \frac{a}{M^2 + a^2} \delta J +  \frac{Q M}{M^2 + a^2} \delta Q \, .
\label{ccv}
\end{equation}

To analyze whether it is possible to produce such a perturbation, let $\Sigma_0$ be an asymptotically flat hypersurface that terminates on the future horizon and extends to spatial infinity. 
\begin{figure}[h]
\begin{center}
\includegraphics[scale=0.6]{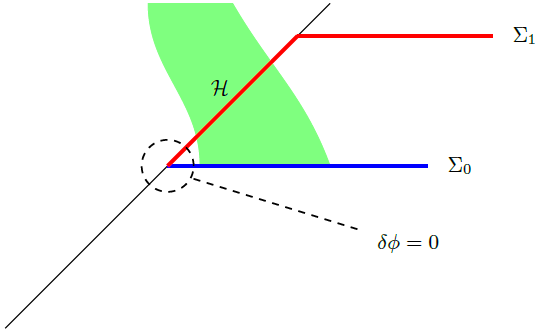}
\caption{Charged matter, occupying the shaded region, falls through the event horizon of an extremal black hole. The perturbed initial data on $\Sigma_0$ vanishes in a neighborhood of the horizon.}
\label{fig:extremal_matter}
\end{center}
\end{figure}
We consider a perturbation $\delta \phi$ whose initial data on $\Sigma_0$ for the fields $\delta g_{ab}$ and $\delta A_{a}$ vanishes in a neighborhood of $\Sigma_0 \cap {\mathcal H}$, as shown in Fig. \ref{fig:extremal_matter}. 
We assume that the matter sources $\delta T_{ab}$ and $\delta j^a$ are nonvanishing only in a compact region of $\Sigma_0$, as shown. 
Physically, this corresponds to considering a perturbation that is induced by bringing matter in from infinity in such a way that the disturbance to the black hole at very early advanced times is negligibly small.
If we evolve the perturbation, in general, some of the matter will go into the black hole and some will go out to infinity or remain in orbit around the black hole. The matter that does not fall into the black hole is of no interest to us. Therefore, we can greatly simplify our analysis by restricting to the case where all of the matter goes into the black hole. Note that this also saves us the trouble of analyzing the motion of bodies outside of the black hole; we do not care about the details of how the matter managed to get into the black hole as long as it does get in.

Thus, we wish to consider a one-parameter family where $\delta T_{ab}$ and $\delta j^a$ are nonvanishing only in a region like the shaded region of Fig. \ref{fig:extremal_matter}. Let $\Sigma$ be a hypersurface like that shown in Fig. \ref{fig:extremal_matter} with the following characteristics: (a) It starts on the future event horizon in a region where the perturbation vanishes. (b) It continues up the future horizon until past the region where the matter sources are nonvanishing. (c) It then becomes spacelike and continues out towards infinity in an asymptotically flat manner. Let $\mathcal{H}$ denote the horizon portion of $\Sigma$, and let $\Sigma_1$ denote the spacelike portion (see Fig. \ref{fig:extremal_matter}) so that
\begin{equation}
\Sigma = \mathcal{H} \cup \Sigma_1 \, .
\end{equation}
We now use (\ref{intid3}) (with $\delta A_B = \delta Q_B = 0$) for this choice of $\Sigma$. The integrand on the right side of (\ref{intid3}) is nonvanishing only on $\mathcal{H}$. Thus, we obtain,
\begin{equation}
	\delta M - \Omega_H \delta J
		= - \int_{\mathcal{H}} \epsilon_{ebcd} \xi_a \delta T^{ae} - \int_{\mathcal{H}} \xi_a A^a \delta ( \epsilon_{ebcd} j^e ) \, .
\end{equation}
Since $\Phi_H = - \xi^a A_a$ is constant on $\mathcal{H}$, we may pull it out of the integral. The integral $\int_{\mathcal{H}} \delta(\epsilon_{ebcd} j^e)$ is just the total flux of electromagnetic charge through the horizon, $\delta Q_{flux}$. Since all of the charge added to the spacetime falls through the horizon, this flux is just equal to the total perturbed charge of the black hole, $\delta Q_{flux}= \delta Q$. Combining these observations yields the following formula relating the perturbed parameters of the black hole spacetime:
\begin{equation}
	\delta M - \Omega_H \delta J - \Phi_H \delta Q
		= - \int_{\mathcal{H}} \epsilon_{ebcd} \xi_a \delta T^{ae} \, . \label{linear_final_formula}
\end{equation}
This result was first derived in \cite{wald2001}. On the horizon, we may write
\begin{equation}
\epsilon_{ebcd} = - 4 k_{[e} \tilde{\epsilon}_{bcd]},
\end{equation}
where $k^a$ is the future-directed normal to the horizon and $\tilde{\epsilon}_{bcd}$ is the corresponding volume element on the horizon. The right side of (\ref{linear_final_formula}) can be written as
\begin{equation}
- \int_{\mathcal{H}} \epsilon_{ebcd} \xi_a \delta T^{ae} =  \int_{\mathcal{H}} \tilde{\epsilon}_{bcd} \delta T^{ae} \xi_a k_e.
\end{equation}
Since $\xi^a \propto k^a$, the right side is non-negative provided only that the non-electromagnetic stress energy tensor $\delta T_{ab}$ satisfies the null energy condition, so that $\delta T_{ab} k^a k^b \geq 0$. 
Thus, (\ref{linear_final_formula}) yields the inequality
\begin{equation}
	\delta M - \Omega_H \delta J - \Phi_H \delta Q
		\geq 0 \, , \label{linear_final_inequality}
\end{equation}
which holds for all perturbations of an extremal Kerr-Newman black hole resulting from charged-matter entering the black hole. 

For a general (not necessarily extremal) Kerr-Newman black hole, we have
\begin{equation}
	\Omega_H
		= \frac{a}{r_+^2 + a^2}
\end{equation}
and
\begin{equation}
	\Phi_H
		= \frac{Q r_+}{r_+^2 + a^2},
\end{equation}
where $r_+$ is the horizon radius
\begin{equation}
	r_+ = M + \sqrt{M^2 - (J/M)^2 - Q^2}.
\end{equation}
For an extremal black hole, we have $r_+ = M$, so (\ref{linear_final_inequality}) yields
\begin{equation}
	\delta M \geq \frac{a}{M^2 + a^2} \delta J + \frac{QM}{M^2 + a^2} \delta Q  \, .
\end{equation}
Thus, (\ref{ccv}) cannot be satisfied, and an extremal black hole cannot be destroyed by dropping/throwing matter into it. This generalizes the results of paper I \cite{wald1974} to arbitrary matter, provided only that the non-electromagnetic contribution to the stress-energy tensor satisfies the null energy condition. This argument that (\ref{linear_final_formula}) implies that one cannot over-charge or over-spin an extremal black hole was previously given in \cite{natario2016}.

\section{Gedanken Experiments to Destroy a Slightly Non-Extremal Black Hole}
\label{sec:nonextremal}

\begin{figure}[h]
\begin{center}
\includegraphics[scale=0.6]{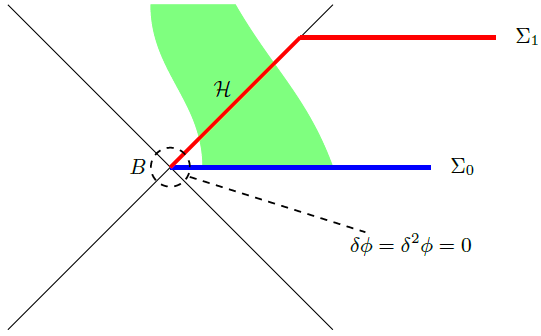}
\caption{A spacetime diagram showing charged matter falling into a black hole as in Fig. 1, but now shown for a non-extremal black hole. The surface $\Sigma_0$ is taken to pass through the bifurcation surface.}
\label{fig:nonextremal_matter}
\end{center}
\end{figure}

In the spirit of Hubeny \cite{hubeny1999}, let us now repeat the gedanken experiment of the previous section starting with a slightly non-extremal Kerr-Newman black hole. The relevant spacetime diagram for this case is shown in Fig. \ref{fig:nonextremal_matter}, where the only significant difference is that $\Sigma_0$ and $\Sigma$ are now taken to terminate at the bifurcation surface, $B$. This does not affect the analysis of the first order perturbation given in the previous section, since the perturbation  is assumed to vanish on the horizon at sufficiently early advanced times. Since we will need to calculate second order effects in this section, we further assume that the second order perturbation also vanishes in a neighborhood of $B$, and that all of the matter sources go into the black hole at second order, so that $\delta^2 T_{ab} = \delta^2 j^a = 0$ on $\Sigma_1$.

An exact repetition of the analysis of the previous section yields
\begin{eqnarray}
\delta M &=& \Omega_H \delta J + \Phi_H \delta Q
		 - \int_{\mathcal{H}} \epsilon_{ebcd} \xi_a \delta T^{ae} \nonumber \\*
	   &\geq&  \Omega_H \delta J + \Phi_H \delta Q \nonumber \\*
	   &=& \frac{a}{r_+^2 + a^2} \delta J + \frac{Q r_+}{r_+^2 + a^2} \delta Q \, .\label{explicit_linear_inequality}
\end{eqnarray}
As already noted in the Introduction for the special case of a nearly extremal Reissner-Nordstrom black hole, this equation admits the possibility of violating (\ref{CC_inequality}). However, as discussed in the Introduction, in order to determine whether violations of (\ref{CC_inequality}) really occur, it is necessary to calculate the second order corrections, $\delta^2 M$, to the mass of the black hole.

In order to proceed further with our analysis of the second order corrections to mass, we will make the following additional assumption:

\medskip

\noindent
\textbf{Additional Assumption}: The (slightly) non-extremal, unperturbed Kerr-Newman black hole we are considering is linearly stable to perturbations, i.e., any source-free\footnote{Our perturbations are, in general, not source-free. However, we will only need to apply this assumption on the late-time surface $\Sigma_1$ sketched in Fig. \ref{fig:nonextremal_matter}, long after all sources have fallen into the black hole.} solution to the linearized Einstein-Maxwell equations approaches a perturbation towards another Kerr-Newman black hole at sufficiently late times.

\medskip

It should be emphasized that this linear stability assumption is entirely compatible with having an instability associated with over-charging or over-spinning the black hole, i.e., we are not assuming what we wish to show. Since we are considering a non-extremal black hole (i.e., $M^2 > (J/M)^2 + Q^2$), a {\em finite} perturbation is required to over-charge or over-spin it. A linear perturbation of a non-extremal black hole always can be scaled down so as to not violate (\ref{CC_inequality}). Thus, the presence of a linear instability of a non-extremal black hole would represent an instability that is independent of over-charging or over-spinning. If a non-extremal black hole were linearly unstable, there would be no need to attempt to over-charge or over-spin it in order to destroy it.

In view of this assumption, we may choose $\Sigma$ in Fig. \ref{fig:nonextremal_matter} so the horizon portion, $\mathcal H$, extends to sufficiently late times that it enters the late time stationary era of the perturbation. We may then take $\Sigma_1$ so that it extends far\footnote{If we wish to take $\Sigma_1$ to extend infinitely far from the black hole, we would have to take it to null infinity rather than spatial infinity.} from the black hole while remaining in the stationary region. The quantities $\delta^2 M$ and $\delta^2 J$ arising in the boundary term (\ref{d2M}) on $\Sigma$ will then have the interpretation of being the second order corrections to the mass and angular momentum of the perturbed black hole\footnote{Note that since mass and angular momentum cannot be radiated away at linear order, we did not need to be careful in our specification of $\Sigma_1$ in our first order analysis in order for $\delta M$ and $\delta J$ to represent the perturbed mass and angular momentum of the final black hole.}.

We now consider our gedanken experiment to destroy the slightly non-extremal black hole. We assume that our first order perturbation has been done optimally (see (\ref{explicit_linear_inequality})), so that 
\begin{equation}
\delta M = \Omega_H \delta J + \Phi_H \delta Q = \frac{a}{r_+^2 + a^2} \delta J + \frac{Q r_+}{r_+^2 + a^2} \delta Q \, .
\label{opt}
\end{equation}
As can be seen from (\ref{explicit_linear_inequality}), this requires vanishing non-electromagnetic energy flux through the horizon, i.e., $\delta T_{ab} k^a k^b = 0$, as should be (nearly) achievable if the matter is lowered (nearly) to the horizon or is (nearly) at a turning point of its orbit just before entering the black hole.

The second order change in mass is given by (\ref{EM_canonical_energy_identity}) with $\delta^2 Q_B = \delta^2 A_B =0$ (since the second order perturbation has been assumed to vanish in a neighborhood of $B$). We have
\begin{equation}
	\delta^2 M - \Omega_H \delta^2 J 
		=\mathcal{E}_{\Sigma}(\phi; \delta \phi)
			- \int_{\mathcal{H}} \iota_{\xi} (\delta \mathbf{E}(\phi) \cdot \delta \phi)
			- \int_{\mathcal{H}} \delta^2 \mathbf{C}_{\xi} \, .
\label{d2M}
\end{equation}
Here, the integrals in the last two terms extend only over $\mathcal H$ rather than over all of $\Sigma = \mathcal{H} \cup \Sigma_1$ because $\delta \mathbf{E}$ and $\delta^2 \mathbf{C}_{\xi}$ vanish on $\Sigma_1$ by the assumption that there are no sources outside the black hole at late times. 

We now evaluate the last two terms appearing on the right side of (\ref{d2M}). 
From (\ref{EM_equations_of_motion}), we have
\begin{equation}
	\left(\iota_{\xi} \left(\delta \mathbf{E}(\phi) \cdot \delta \phi \right)\right)_{abc}
		= - \xi^d \epsilon_{dabc} \left[ \frac{1}{2} \delta T^{ef} \delta g_{ef} + \delta j^e \delta A_e \right].
\end{equation}
Since $\xi^a$ is tangent to the horizon, the pullback to $\mathcal{H}$ of this term vanishes, so it does not contribute to (\ref{d2M}).
From (\ref{EM_constraints}), we have 
\begin{equation}
\left(\delta^2 \mathbf{C}_{\xi} \right)_{abc} = \delta^2 \left(\epsilon_{eabc} {T_d}^e \xi^d \right) + \delta^2 \left(\epsilon_{eabc} A_d j^e \xi^d \right).
\end{equation}
Using our gauge condition $\xi^a \delta A_a= 0$ on $\mathcal H$ (see (\ref{Maxwell_gauge_condition}) and the discussion of subsection \ref{subsec:gauge}), we see that on $\mathcal H$, the second term is 
\begin{equation}
\delta^2 \left(\epsilon_{eabc} A_d j^e \xi^d \right) = - \Phi_H \delta^2 \left( \epsilon_{eabc}  j^e \right),
\end{equation}
and therefore
\begin{equation}
	\delta^2 \left[ \int_{\mathcal{H}} \xi_a A^a \epsilon_{ebcd} j^e \right]
		= - \Phi_H \delta^2 Q_{flux} = - \Phi_H \delta^2 Q,
\end{equation}
where $\delta^2 Q$ is the second-order change in charge of the black hole. On the other hand, using our assumption that the first order process was done optimally and thus there was vanishing non-electromagnetic stress-energy flux through the horizon at first order, we have
\begin{equation}
\delta^2 \left(\epsilon_{eabc} {T_d}^e \xi^d \right) = \epsilon_{eabc}  \xi^d \delta^2 {T_d}^e.
\end{equation}
Putting this together, we obtain
\begin{equation}
	\delta^2 M - \Omega_H \delta^2 J - \Phi_H \delta^2 Q
		= \mathcal{E}_{\Sigma}(\phi; \delta \phi)
			-  \int_{\mathcal{H}}  \xi^a  \epsilon_{ebcd}  \delta^2 {T_a}^e  \, .
\end{equation}
The last term in this equation is positive provided that the non-electromagnetic stress-energy tensor satisfies the null energy condition.

It remains to compute the canonical energy $\mathcal{E}_ \Sigma (\phi; \delta \phi)$. Since  
$\Sigma = \mathcal{H} \cup \Sigma_1$, we have
\begin{equation}
	\mathcal{E}_{\Sigma}(\phi; \delta \phi)
		= \int_{\mathcal{H}} \bm{\omega}(\phi, \delta \phi, \mathscr{L}_{\xi} \delta \phi) + \int_{\Sigma_1} \bm{\omega}(\phi, \delta \phi, \mathscr{L}_{\xi} \delta \phi) \, . \label{canonical_energy_integral}
\end{equation}
Let us calculate first calculate the horizon contribution. We have
\begin{equation}
\int_{\mathcal{H}} \bm{\omega}=  \int_{\mathcal{H}} {\bm{\omega}}^{GR} + \int_{\mathcal{H}} {\bm{\omega}}^{EM} \, ,
\end{equation}
where the gravitational and electromagnetic parts, ${\bm{\omega}}^{GR} $ and ${\bm{\omega}}^{EM}$, are given, respectively, by (\ref{GR_symplectic_current}) and (\ref{electromagnetic_symplectic_current}) above.
The integral over $\mathcal{H}$ of the gravitational part of the canonical energy density was computed in \cite{wald2012}, and is given by\footnote{Eq.(\ref{GR_CE_density}) assumes that $\delta \Theta = 0$ on $\mathcal H$ (see \cite{wald2012}). This condition can be imposed in the present case because we assumed that the first order process was done optimally [see (\ref{opt})], so $\delta T_{ab} k^a k^b = 0$.}
\begin{eqnarray}
	\int_{\mathcal{H}} \bm{\omega}^{GR}(g; \delta g,\mathscr{L}_{\xi} \delta g)
		& = & \frac{1}{4\pi} \int_{\mathcal{H}} (\xi^a \nabla_a u) \delta \sigma_{bc} \delta \sigma^{bc} \bm{\epsilon} \nonumber \\*
		&& + \frac{1}{16\pi} \int_S (\xi^a \nabla_a u) \delta g^{bc} \delta \sigma_{bc} \bm{\epsilon} \quad\quad \label{GR_CE_density}
\end{eqnarray}
where $\delta \sigma_{ab}$ denotes the perturbed shear of the horizon generators, $u$ is an affine parameter along the future horizon, and $S = {\mathcal{H} \cap \Sigma_1}$ is the $2$-surface formed by the intersection of $\mathcal{H}$ and $\Sigma_1$. By our additional assumption above, the perturbation is physically stationary at $S$, so $\delta \sigma_{ab} = 0$ on $S$. Thus, we obtain
\begin{equation}
	\int_{\mathcal{H}} \bm{\omega}^{GR}(\phi; \delta \phi,\mathscr{L}_{\xi} \delta \phi)
		 =  \frac{1}{4\pi} \int_{\mathcal{H}} (\xi^a \nabla_a u) \delta \sigma_{bc} \delta \sigma^{bc} \bm{\epsilon} \geq 0 \, . \label{GR_CE_flux_matter}
\end{equation}
We may interpret this horizon flux contribution from ${\bm{\omega}}^{GR}$ as representing the total flux of gravitational wave energy into the black hole.

Next, we calculate the horizon flux contribution from ${\bm{\omega}}^{EM}$. From (\ref{electromagnetic_symplectic_current}), we have
\begin{widetext}
\begin{equation}
	(\omega^{EM})_{abc}(\phi; \delta \phi, \mathscr{L}_{\xi} \phi)
		= \frac{1}{4\pi} \epsilon_{dabc} \left[ \delta A_e \mathscr{L}_{\xi} \delta F^{de}
		- \delta F^{de} \mathscr{L}_{\xi} \delta A_e \right]
		+ \frac{1}{4\pi} \left[ (\mathscr{L}_{\xi} \delta \epsilon_{dabc}) F^{de}\delta A_e
		 - (\delta \epsilon_{dabc}) F^{de} \mathscr{L}_{\xi} \delta A_e\right].
		\label{electromagnetic_CE_density}
\end{equation}
\end{widetext}
The last two terms on the right side of this equation involve the background electromagnetic field strength $F^{ab}$. However, by (\ref{HEM}) together with our gauge condition $\xi^a \delta A_a = 0$ on $\mathcal H$, it can be seen that the last two terms in (\ref{electromagnetic_CE_density}) vanish. The first term in (\ref{electromagnetic_CE_density}) can be written as
\begin{equation}
\epsilon_{dabc} \delta A_e \mathscr{L}_{\xi} \delta F^{de} = \mathscr{L}_{\xi} \left[ \epsilon_{dabc} \delta A_e \delta F^{de} \right] -  \epsilon_{dabc}  \delta F^{de} \mathscr{L}_{\xi} \delta A_e \, .
\label{term1}
\end{equation}
When pulled back to $\mathcal H$, $\epsilon_{dabc} \delta A_e \delta F^{de}$ is a $3$-form $\bm{\eta}$, on a $3$-dimensional surface, so when pulled back to $\mathcal H$, we have
\begin{equation}
\mathscr{L}_{\xi} \bm{\eta} = \iota_\xi d \bm{\eta} + d \left(\iota_\xi  \bm{\eta} \right) = d \left(\iota_\xi  \bm{\eta} \right),
\end{equation}
where the pullback of $\iota_{\xi} d \bm{\eta}$ vanishes since $\xi^a$ is tangent to $\mathcal{H}.$ Thus, the integral over $\mathcal H$ of the first term on the right side of (\ref{term1}) will merely contribute a boundary term at $S = {\mathcal{H} \cap \Sigma_1}$. However, since the perturbation is assumed to be stationary at $S$, the electromagnetic energy flux must vanish there, so $\delta F_{ab}$ must be of the form (\ref{HEM}). Using this fact together with our gauge condition $\xi^a \delta A_a = 0$ on $\mathcal H$, it can be seen that this boundary term vanishes. 
Finally, the second term on the right side of (\ref{term1}) combines with the second term of (\ref{electromagnetic_CE_density}). This term can be further simplified by noting that 
\begin{equation}
\mathscr{L}_{\xi} \delta \bm{A} = \iota_\xi d \delta \bm{A} + d \left( \iota_\xi \delta \bm{A} \right). \label{maxwell_lie_derivative}
\end{equation}
Under our gauge condition $\xi^a \delta A_a|_{\mathcal{H}} = 0$, the second term of (\ref{maxwell_lie_derivative}) is normal to the horizon, and hence proportional to the horizon normal $k^a$. By the antisymmetry of $\delta F_{ab}$, $\delta F^{ab} k_b$ is orthogonal to $k^a$ and hence tangent to the horizon. As this term only appears in (\ref{electromagnetic_CE_density}) when contracted into the volume element on the horizon, it makes no contribution to the canonical energy integral. Putting everything together, we find that 
\begin{equation}
\int_{\mathcal{H}} \bm{\omega}^{EM}(\phi; \delta \phi,\mathscr{L}_{\xi} \delta \phi) = - \frac{1}{2 \pi} \int_\mathcal{H} \epsilon_{dabc} \xi^e \delta F^{df} \delta F_{ef} \, .
\label{emflux}
\end{equation}
The right side of this equation is nonnegative and can be interpreted as the total flux of electromagnetic energy into the black hole.

All that remains now is to calculate the contribution to canonical energy from $\Sigma_1$
\begin{equation}
	\mathcal{E}_{\Sigma_1}(\phi; \delta \phi)
		= \int_{\Sigma_1} \bm{\omega}(\phi, \delta \phi, \mathscr{L}_{\xi} \delta \phi) \, . 
\label{E1}
\end{equation}
Since we have assumed that the perturbation is stationary on $\Sigma_1$, it might be thought that 
$\mathscr{L}_{\xi} \delta \phi = 0$ on $\Sigma_1$ and thus this contribution to the canonical energy vanishes. However, this is not the case because our conditions $\delta \xi^a = 0$ as well as our gauge condition $\xi^a \delta A_a = 0$ on $\mathcal{H}$ preclude our writing the perturbation in a gauge where $\mathscr{L}_{\xi} \delta g_{ab} = 0$ and $\mathscr{L}_{\xi} \delta A_a =0$; see \cite{wald2012} for further discussion. Nevertheless, we can calculate $\mathcal{E}_{\Sigma_1}(\phi; \delta \phi)$ as follows. First, since, by assumption, $\delta \phi$ is equal to a perturbation $\delta \phi^{KN}$ to another Kerr-Newman black hole on $\Sigma_1$, we obviously may replace $\delta \phi$ by $\delta \phi^{KN}$ (written in our gauge) on the right side of (\ref{E1})
\begin{equation}
	\mathcal{E}_{\Sigma_1}(\phi; \delta \phi) = \mathcal{E}_{\Sigma_1}(\phi; \delta \phi^{KN})
		= \int_{\Sigma_1} \bm{\omega}(\phi, \delta \phi^{KN}, \mathscr{L}_{\xi} \delta \phi^{KN}) \, . 
\label{E2}
\end{equation}
However, as can be seen from our analysis above, $\delta \phi^{KN}$ has no flux of canonical energy through $\mathcal H$, i.e., there is no flux of gravitational or electromagnetic energy through the horizon for a Kerr-Newman perturbation. Thus, we may replace $\Sigma_1$ by $\Sigma$ in (\ref{E2}). Finally, we may evaluate $\mathcal{E}_{\Sigma}(\phi; \delta \phi^{KN})$ using (\ref{EM_canonical_energy_identity}). 
Consider the one-parameter family, $\phi^{KN}(\alpha)$, where each field configuration in the family is a Kerr-Newman black hole with parameters given by
\begin{eqnarray}
	M^{KN}(\alpha)
		& = & M + \alpha \delta M, \label{KN_mass_perturbation} \\
	Q^{KN}(\alpha)
		& = & Q + \alpha \delta Q, \label{KN_charge_perturbation} \\
	J^{KN}(\alpha)
		& = & J + \alpha \delta J, \label{KN_angular_momentum_perturbation}
\end{eqnarray}
where $\delta M,$ $\delta Q,$ and $\delta J$ are chosen to agree with the corresponding values for  our first-order perturbation $\phi(\lambda)$. Then, for this family, we have $\delta^2 M = \delta^2 J = \delta^2 Q_B = 0$, as well as $\delta {\bm E} = \delta^2 {\bm C}_\xi = 0$. Thus, we obtain
\begin{equation}
	\mathcal{E}_{\Sigma}(\phi; \delta \phi^{KN})
		= - \frac{\kappa}{8 \pi} \delta^2 A_B^{KN}.
\end{equation}
where $\delta^2 A_B^{KN}$ denotes the second order change in the area of the horizon for the one-parameter family (\ref{KN_mass_perturbation})-(\ref{KN_angular_momentum_perturbation}).
This quantity can be evaluated by taking two variations of the area formula $A_B = 4 \pi (r_+^2 + (J/M)^2),$ and is given explicitly as follows:
\begin{widetext}
\begin{eqnarray}
	\delta^2 A_B^{KN}
		& = - \frac{8 \pi}{M^8 \epsilon^3}
			& \left[ (\delta M)^2 \left(J^4 + (2 + \epsilon^2) J^2 M^4 - M^8 (1 + \epsilon) (-1 + \epsilon + 2 \epsilon^2) \right) \right. \nonumber \\*
		&& \left. 	+ (\delta Q)^2 \left(M^6 Q^2 + M^8 (1 + \epsilon) \epsilon^2\right) + (\delta J)^2 \left(J^2 M^2 + M^6 \epsilon^2\right) \right. \nonumber \\*
		&& \left.	+ \delta M \delta J \left(-2 J^3 M - 2 J M^5 (1 + \epsilon^2)\right) + \delta J \delta Q \left(2 J M^4 Q\right) \right. \nonumber \\*
		&& \left.	+ \delta M \delta Q \left(- 2 J^2 M^3  Q+ 2 M^7 Q (-1 + \epsilon^2)\right) \right] \, . \label{quadratic_area_variation}
\end{eqnarray}
\end{widetext}
Here we have introduced the parameter
\begin{equation}
\epsilon = r_+ / M - 1 = \frac{\sqrt{M^2 - Q^2 - (J/M)^2}}{M}
\label{epsilon}
\end{equation} 
(thereby generalizing (\ref{eps}) to the case where the black hole is rotating as well as charged) in order that we can keep better track of the extremal limit, $\epsilon \rightarrow 0$. However, we have not assumed that $\epsilon$ is small in (\ref{quadratic_area_variation}).

We have now computed all of the terms appearing in (\ref{d2M}). Using the positivity of the gravitational, electromagnetic, and non-electromagnetic stress-energy fluxes through the horizon, we have thereby derived the following inequality involving the second order change of the mass of the black hole
\begin{equation}
	\delta^2 M - \Omega_H \delta^2 J - \Phi_H \delta^2 Q
		\geq - \frac{\kappa}{8 \pi} \delta^2 A_B^{KN}. \label{quadratic_penultimate_bound}
\end{equation}
The surface gravity of a Kerr-Newman black hole is given by
\begin{equation}
	\kappa
		= \frac{M^3}{M^4 (1+ \epsilon)^2 + J^2} \epsilon.
\label{kappa}
\end{equation}
Expanding the right side of (\ref{quadratic_penultimate_bound}) to lowest order in $\epsilon$, we obtain
\begin{widetext}
\begin{equation}
	\delta^2 M - \Omega_H \delta^2 J - \Phi_H \delta^2 Q
		\geq \frac{M}{(M^4 + J^2)^2} \left[ M^4 (\delta J)^2 + (M^6 + J^2 Q^2 + M^2 J^2) (\delta Q)^2
							 - 2 J M^2 Q \delta J \delta Q  \right] + O(\epsilon),
		\label{quadratic_final_bound}
\end{equation}
\end{widetext}
where we have used $\delta M = \Omega_H \delta J + \Phi_H \delta Q$ (see (\ref{opt})) to eliminate 
$\delta M$ from the expression.  

We now show that this inequality is precisely what is needed to show that gedanken experiments of the Hubeny type can never succeed in over-charging or over-spinning the black hole. Consider a one-parameter family, $\phi(\lambda)$, of the type we have been considering, where $\phi(0)$ is a nearly extremal Kerr-Newman black hole, $\epsilon \ll 1$. Define
\begin{equation}
f(\lambda) = M(\lambda)^2 - Q(\lambda)^2 - J(\lambda)^2/M(\lambda)^2
\end{equation}
Then, to second order in $\lambda$, we have
\begin{widetext}
\begin{eqnarray}
	f(\lambda)
		&=& \left(M^2 - Q^2 - \frac{J^2}{M^2} \right)
			+ 2 \lambda \left( \frac{M^4 + J^2}{M^3} \delta M - \frac{J}{M^2} \delta J - Q \delta Q \right) \nonumber \\*
		&& + \lambda^2 \left[ \left(\frac{J^2 + M^4}{M^3} \right) \delta^2 M - \frac{J}{M^2} \delta^2 J  - Q \delta^2 Q + \frac{4 J}{M^3} \delta J \delta M \right. \nonumber \\*
		&& \left. - \frac{1}{M^2} (\delta J)^2 + \left( \frac{M^4 - 3 J^2}{M^4} \right) (\delta M)^2 - (\delta Q)^2 \right].
		\label{CC_quantity_second_order}
\end{eqnarray}
\end{widetext}
We wish to know if, for small, $\lambda$, we can make $f < 0$. If we took into account only effects linear in $\lambda$, the inequality (\ref{explicit_linear_inequality}) would constrain $f$ by 
\begin{eqnarray}
	f(\lambda)
		 & \geq &  M^2 \epsilon^2 + \frac{2}{M^4 + J^2} \left( (J^2 - M^4) Q \delta Q - 2 J M^2 \delta J \right)\lambda \epsilon \nonumber \\*
		&& + O(\lambda^2, \epsilon^3, \epsilon^2 \lambda) \, .
\end{eqnarray}
If the $O(\lambda^2)$ term and the higher order terms are neglected, then it is easy to see that it is possible to make $f(\lambda) < 0$, suggesting that the black hole could be over-charged or over-spun. However, when our calculation of the $O(\lambda^2)$ term given by inequality (\ref{quadratic_final_bound}) is taken into account, we have shown that for an optimal first-order process with $\delta M = \Omega_H \delta J + \Phi_H \delta Q$, we have 
\begin{eqnarray}
	f(\lambda)
		& \geq & M^2 \epsilon^2
			+ \frac{2}{M^4 + J^2} \left( (J^2 - M^4) Q \delta Q - 2 J M^2 \delta J \right)\lambda \epsilon \nonumber \\*
		&&	+ \frac{1}{M^2 (M^4 + J^2)^2} ((J^2 - M^4) Q \delta Q - 2 J M^2 \delta J)^2 \lambda^2 \nonumber \\*
		&& + O(\lambda^3, \epsilon^3, \epsilon^2 \lambda, \epsilon \lambda^2),
\end{eqnarray}
This expression can be rewritten as a perfect square,
\begin{eqnarray}
	f(\lambda)
		& \geq & \left( \frac{(J^2 - M^4) Q \delta Q - 2 J M^2 \delta J}{M (M^4 + J^2)} \lambda + M \epsilon \right)^2 \nonumber \\*
		&& + O(\lambda^3, \dots).
		\label{final_CC_result}
\end{eqnarray}
Thus, $f \geq 0$, and no violations of (\ref{CC_inequality}) can occur.

\section{Discussion}
\label{sec:discussion}

\begin{figure}[h]
\begin{center}
\includegraphics[scale=0.6]{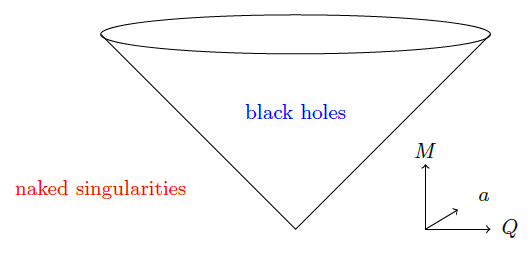}
\caption{The parameter space of Kerr-Newman black holes.}
\label{fig:KN}
\end{center}
\end{figure}

The Kerr-Newman parameter space $(M, Q, a=J/M)$ is shown in Fig. \ref{fig:KN}. In this parameter space, black holes lie within the ``future light cone'' $M > 0$, $M^2 - Q^2 - a^2 \geq 0$. Kerr-Newman solutions outside this cone correspond to naked singularities.
Extremal black holes live on the boundary of the cone,  $M = \sqrt{Q^2 + a^2}$. The gedanken experiments to destroy an extremal black hole discussed in section \ref{sec:extremal} correspond to analyzing whether, starting at the boundary, one can perturb the spacetime so as to move outside the cone. The gedanken experiments to destroy a slightly non-extremal black hole discussed in section \ref{sec:nonextremal} correspond to analyzing whether one can move out of the cone starting near (but not on) the boundary of the cone.

\begin{figure}[h]
\begin{center}
\includegraphics[scale=0.6]{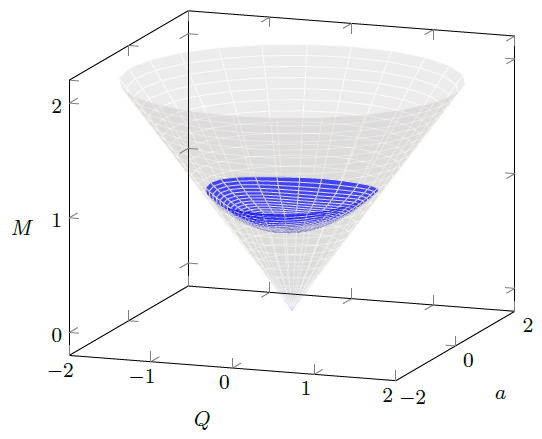}
\caption{A surface of constant area for Kerr-Newman black holes.}
\label{fig:area}
\end{center}
\end{figure}

Within this cone, one can draw surfaces of constant area for the Kerr-Newman black holes. One such surface is shown in Fig. \ref{fig:area}. It is important to note that the surfaces of constant area meet the boundary tangentially.

To linear order, the change in the parameters $(M,Q,a)$ resulting from dropping matter into a Kerr-Newman black hole corresponds to a tangent vector in parameter space.
Equation (\ref{explicit_linear_inequality}) shows precisely that for an arbitrary Kerr-Newman black hole, to linear order, any perturbation resulting from matter entering a black hole cannot decrease the area of the black hole\footnote{This result was first obtained for particle matter by Christodoulou \cite{christodoulou1970}.}. Thus, the tangent to the surface of constant area provides a lower bound to the slope of any tangent vector representing a physically achievable perturbation. In particular, for an extremal black hole, the best one can do is move tangentially to the cone. Thus, as we found in section \ref{sec:extremal}, to first order it is impossible to escape from the cone into the naked singularity region of parameter space starting at the boundary of the cone.

\begin{figure}[h]
\begin{center}
\includegraphics[scale=0.6]{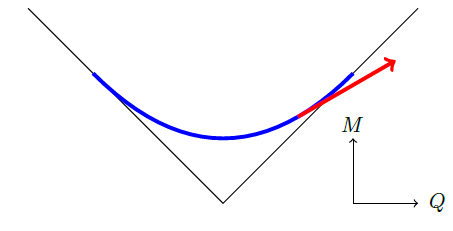}
\caption{The tangent to a curve of constant area for a slightly non-extremal Reissner-Nordstrom black hole.}
\label{fig:hub}
\end{center}
\end{figure}

The Hubeny argument for possibly escaping from the cone is illustrated in Fig. \ref{fig:hub}. For simplicity in the drawing, we have set $J=0$ and thus show only the parameter space of Reissner-Nordstrom solutions.
As is illustrated in this figure, except at the boundary, the tangent to the curve of constant area has a slope strictly less than one. Thus, a straight line tangent to such a curve will exit the cone. This means that if the linear approximation were valid for a finite perturbation, it would be possible to add charged matter to a slightly non-extremal Reissner-Nordstrom black hole so as to over-charge the black hole, as originally argued by Hubeny. 

\begin{figure}[h]
\begin{center}
\includegraphics[scale=0.6]{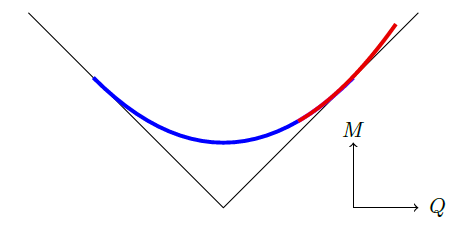}
\caption{The quadratic approximation to the curve of final state parameters obtained by adding charged matter to a slightly non-extremal Reissner-Nordstrom black hole.}
\label{fig:hub2}
\end{center}
\end{figure}

However, our work shows that at second order, there are corrections to the straight line, as illustrated in Fig. \ref{fig:hub2}. 
Consider a one-parameter family of solutions corresponding to adding charged matter to the black hole. As we have noted above, the curve representing the final state parameters has a tangent whose slope is bounded below by the tangent to the curve of constant area. In addition, however, if its slope is the minimum possible, we have proven in section \ref{sec:nonextremal} that the second derivative of the curve must be greater than the second derivative of the curve of constant area. The quadratic approximation to this curve thus coincides with the curve of constant area and does not exit the cone. The linear approximation is simply not an adequate approximation. Second order effects do not allow one to exit from the cone.

Finally, it is worth noting that there is a discontinuity in our lower bound on $\delta^2 M$ in the extremal limit. Consider, for simplicity, the case of adding charged matter with no angular momentum to a Reissner-Nordstrom black hole, so $J= \delta J = \delta^2 J = 0$. Without loss of generality, we also may take $\delta^2 Q = 0$. Then, for $\epsilon > 0$, for an optimal perturbation with $\delta M = \Phi_H \delta Q$, it follows from (\ref{quadratic_final_bound}) that 
\begin{equation}
	\delta^2 M \geq \frac{(\delta Q)^2}{M} + O(\epsilon) \, .
\label{extlim}
\end{equation}
Thus, as $\epsilon \to 0$, the right side approaches $(\delta Q)^2/M$. Now consider adding charged matter to an exactly extremal black hole, $\epsilon = 0$. As shown in section \ref{sec:extremal}, the optimal perturbation satisfies $\delta M = \Phi_H \delta Q = \delta Q$, so optimally, the perturbation moves one tangent to the cone. However, the derivation of (\ref{quadratic_final_bound}) does {\em not} apply to this case---even if we assume that the linearized perturbation becomes stationary at late times---because our evaluation of ${\mathcal E}_{\Sigma_1}$ is valid only for non-extremal black holes. Nevertheless, if the perturbation {\em decreases} the charge of the black hole (i.e., if $\delta Q$ has sign opposite that of $Q$) then one would expect that $\delta^2 M \geq (\delta Q)^2/M$, so that, optimally, at second order the area of the black hole will remain constant. On the other hand, if $\delta Q$ {\em increases} the charge, then there is no reason why this bound need be satisfied since the area of the black hole will increase in any case. Our expectation is that 
\begin{equation}
	\delta^2 M \geq 0 \, ,
\label{extlb}
\end{equation}
so that, optimally, the black hole will remain extremal at second order.
Indeed, the explicit example of adding a charged shell of matter shows that the lower bound (\ref{extlb}) can, in fact, be achieved. Thus, there is a discontinuity between (\ref{extlim}) and (\ref{extlb}) when $\epsilon \to 0$. It would be interesting to derive (\ref{extlb}) from first principles and to see if it is related to other discontinuous behavior as $\epsilon \to 0$, such as the Aretakis instability \cite{aretakis}.

\begin{acknowledgments}
We thank Kartik Prabhu for helpful discussions on the gauge invariance of canonical energy. We also thank Hongbao Zhang for numerous comments on the original version of this manuscript, including pointing out several errors. This research was supported in part by NSF grant PHY 15-05124 to the University of Chicago.
\end{acknowledgments}

\appendix*

\section{Self-Force Energy and Finite Size Effects}

The second-order correction to the mass of a black hole given by equation (\ref{quadratic_final_bound}) gives a lower bound on the energy of any matter that enters a black hole that is valid to quadratic order in the charge and angular momentum of the body. Since particle-like matter in general relativity must be described as a limiting case of general continuum matter (see \cite{gralla_wald_1} and \cite{gralla_wald_2}), this formula applies to particle matter as well. At second order, self-force effects contribute to the energy of a particle. In addition, at second order, a charged body will have an electromagnetic self-energy that diverges when the size of the body is taken to zero, so the size of the body must be finite. However, the finite size of the body may prevent one from lowering the body all the way to the horizon. Our bound (\ref{quadratic_penultimate_bound}) must implicitly take into account all of these effects. The purpose of this Appendix is to show explicitly that this is the case for the special case of a charged, particle-like body that enters an uncharged, non-extremal Kerr black hole along the black hole's symmetry axis. The self-force effects in this case were previously calculated by Leaute and Linet \cite{linet1982}, while self-energy and finite size effects in this case were previously obtained by Hod \cite{hod2002}.

It is particularly easy to evaluate our lower bound on $\delta^2 M$ for the case of a charged body entering a Kerr black hole along the symmetry axis, since $Q=0$ and $\delta J = \delta^2 J = 0$. An optimal process therefore has $\delta M = 0$ at first order. Thus, (\ref{quadratic_area_variation}) reduces to\footnote{Note that this is an exact expression, i.e., we have not assumed that $\epsilon$ is small.}
\begin{equation}
	\delta^2 A_B^{KN} = - \frac{8 \pi}{\epsilon} (1 + \epsilon) (\delta Q)^2
\end{equation}
Hence, (\ref{quadratic_penultimate_bound}) yields
\begin{equation}
	\delta^2 M \geq - \frac{1}{8 \pi} \kappa \delta^2 A_B^{KN} = \frac{r_+}{r_+^2 + a^2} (\delta Q)^2
	\label{sfe}
\end{equation}
where we have used the expression (\ref{kappa}) for $\kappa$ and have used (\ref{epsilon}) to replace $\epsilon$ by $r_+$. Since $Q=0$, we have $r_+^2 + a^2 = 2 M r_+$, and so (\ref{sfe}) may be written as
\begin{equation}
	\delta^2 M \geq  \frac{1}{2 M} (\delta Q)^2.
	\label{sfe2}
\end{equation}
Taking into account the Taylor coefficient of 1/2, this means that any charged matter with no angular momentum that enters an uncharged black hole must carry an energy
\begin{equation}
	E \geq \frac{1}{4 M} (\delta Q)^2.
	\label{sfe3}
\end{equation}
into the black hole. This bound holds for any Kerr black hole with $a < M$.

On the other hand, Leaute and Linet's expression \cite{linet1982} for the (proper, locally measured) self-force on a charged particle on the symmetry axis of Kerr is repulsive and has magnitude
\begin{equation}
	f(r)
		= \frac{M r}{(r^2 + a^2)^2} (\delta Q)^2.
\end{equation}
The force exerted at infinity when lowering the charged body is reduced from this by the redshift factor $(-g_{tt})^{1/2}$ (see, e.g., \cite{wald1982}). However, the infinitesimal proper distance traversed when lowering is given by $dl = (g_{rr})^{1/2} dr$. The factors $(-g_{tt})^{1/2}$ and $(g_{rr})^{1/2}$ cancel on the symmetry axis of Kerr.
Thus, we find that the work done at infinity in overcoming the self-force when lowering the charge from infinity to the horizon is
\begin{equation}
	E_{SF}
		= \int_{r_+}^{\infty} f(r) dr
		= \frac{M}{2 (r_+^2 + a^2)} (\delta Q)^2 \, .
	\label{sf_contribution}
\end{equation}
Note that $E_{SF} < E_{min}$ for a nonextremal black hole, with $E_{min}$ given by the right side of (\ref{sfe3}).

However, the self-force expression is only valid for a small body that is roughly spherical in shape. 
For such a body, there will be potentially important self-energy and finite size effects, which can be calculated as follows. For a charged spherical body of radius $R$ and charge $\delta Q$, the electromagnetic contribution to the rest mass of the body is minimized for a thin shell and is given by
\begin{equation}
	m_{EM}
		= \frac{1}{2} \frac{(\delta Q)^2}{R} \, .
\end{equation}
If the body is dropped into the black hole from a proper distance $l$ from the horizon, its electromagnetic self-energy will contribute an energy
\begin{equation}
	E_{self} = m_{EM} V(l) 
\end{equation}
to the black hole, where $V(l)$ is the redshift factor at the dropping point. However, near the black hole, we have
\begin{equation}
	V(l) = \kappa l,
\end{equation}
where $\kappa$ is the surface gravity of the black hole. Since we must have $l \geq R$, we obtain
\begin{equation}
	E_{self}
		\geq \frac{\kappa}{2} (\delta Q)^2.
\end{equation}
Substituting for $\kappa$ from (\ref{kappa}) and adding these two contributions yields a minimal total added energy of
\begin{equation}
 E_{self} + E_{SF}
		= \frac{(\delta Q)^2}{4 M} \, ,
\end{equation}
in exact agreement\footnote{For a nearly extremal Kerr black hole, this is sufficient to prevent over-extremizing the black hole, as previously found by Hod \cite{hod2002}.} with (\ref{sfe3}). Thus, we see explicitly in this example how our general bound 
(\ref{sfe3}) incorporates both self-force effects and self-energy/finite size effects.

One could attempt to evade our bound by making $E_{self}$ smaller by choosing, instead of a small spherical shell, a body that has radial extent much smaller than its angular extent. Such a body could be lowered arbitrarily close to the black hole without making its self-energy arbitrarily large. However, choosing such a shape for the body would result in other second-order corrections to the energy (such as self-repulsion effects) that would inevitably have to reproduce our bound (\ref{sfe3}). As an extreme example of this, one can consider a thin spherical shell of charge collapsing around a Schwarzchild black hole, which experiences a large self-repulsion but for which the (redshifted) electromagnetic self-energy can be made exactly zero. Using the methods of Boulware \cite{boulware1973}, it is straightforward to show that such a shell still adds a minimal energy of $(\delta Q)^2 / 4 M$ to the black hole. This illustrates, again, that our bound automatically takes all effects on energy into account.

\providecommand{\noopsort}[1]{}\providecommand{\singleletter}[1]{#1}%


\begin{thebibliography}{31}%
\makeatletter
\providecommand \@ifxundefined [1]{%
 \@ifx{#1\undefined}
}%
\providecommand \@ifnum [1]{%
 \ifnum #1\expandafter \@firstoftwo
 \else \expandafter \@secondoftwo
 \fi
}%
\providecommand \@ifx [1]{%
 \ifx #1\expandafter \@firstoftwo
 \else \expandafter \@secondoftwo
 \fi
}%
\providecommand \natexlab [1]{#1}%
\providecommand \enquote  [1]{``#1''}%
\providecommand \bibnamefont  [1]{#1}%
\providecommand \bibfnamefont [1]{#1}%
\providecommand \citenamefont [1]{#1}%
\providecommand \href@noop [0]{\@secondoftwo}%
\providecommand \href [0]{\begingroup \@sanitize@url \@href}%
\providecommand \@href[1]{\@@startlink{#1}\@@href}%
\providecommand \@@href[1]{\endgroup#1\@@endlink}%
\providecommand \@sanitize@url [0]{\catcode `\\12\catcode `\$12\catcode
  `\&12\catcode `\#12\catcode `\^12\catcode `\_12\catcode `\%12\relax}%
\providecommand \@@startlink[1]{}%
\providecommand \@@endlink[0]{}%
\providecommand \url  [0]{\begingroup\@sanitize@url \@url }%
\providecommand \@url [1]{\endgroup\@href {#1}{\urlprefix }}%
\providecommand \urlprefix  [0]{URL }%
\providecommand \Eprint [0]{\href }%
\providecommand \doibase [0]{http://dx.doi.org/}%
\providecommand \selectlanguage [0]{\@gobble}%
\providecommand \bibinfo  [0]{\@secondoftwo}%
\providecommand \bibfield  [0]{\@secondoftwo}%
\providecommand \translation [1]{[#1]}%
\providecommand \BibitemOpen [0]{}%
\providecommand \bibitemStop [0]{}%
\providecommand \bibitemNoStop [0]{.\EOS\space}%
\providecommand \EOS [0]{\spacefactor3000\relax}%
\providecommand \BibitemShut  [1]{\csname bibitem#1\endcsname}%
\let\auto@bib@innerbib\@empty
\bibitem [{\citenamefont {Penrose}(1969)}]{penrose1969}%
  \BibitemOpen
  \bibfield  {author} {\bibinfo {author} {\bibfnamefont {R.}~\bibnamefont
  {Penrose}},\ }\href@noop {} {\bibfield  {journal} {\bibinfo  {journal} {Riv.
  Nuovo Cimento}\ }\textbf {\bibinfo {volume} {1}},\ \bibinfo {pages} {252}
  (\bibinfo {year} {1969})}\BibitemShut {NoStop}%
\bibitem [{\citenamefont {Wald}(1997)}]{wald1997}%
  \BibitemOpen
  \bibfield  {author} {\bibinfo {author} {\bibfnamefont {R.~M.}\ \bibnamefont
  {Wald}},\ }\href@noop {} {} (\bibinfo {year} {1997}),\ \Eprint
  {http://arxiv.org/abs/gr-qc/9710068} {arXiv:gr-qc/9710068} \BibitemShut
  {NoStop}%
\bibitem [{\citenamefont {Wald}(1974)}]{wald1974}%
  \BibitemOpen
  \bibfield  {author} {\bibinfo {author} {\bibfnamefont {R.~M.}\ \bibnamefont
  {Wald}},\ }\href@noop {} {\bibfield  {journal} {\bibinfo  {journal} {Annals
  of Physics}\ }\textbf {\bibinfo {volume} {82}},\ \bibinfo {pages} {548}
  (\bibinfo {year} {1974})}\BibitemShut {NoStop}%
\bibitem [{\citenamefont {Hubeny}(1999)}]{hubeny1999}%
  \BibitemOpen
  \bibfield  {author} {\bibinfo {author} {\bibfnamefont {V.~E.}\ \bibnamefont
  {Hubeny}},\ }\href@noop {} {\bibfield  {journal} {\bibinfo  {journal} {Phys.
  Rev. D}\ }\textbf {\bibinfo {volume} {59}},\ \bibinfo {pages} {064013}
  (\bibinfo {year} {1999})}\BibitemShut {NoStop}%
\bibitem [{\citenamefont {de~Felice}\ and\ \citenamefont
  {Yu}(2001)}]{felice2001}%
  \BibitemOpen
  \bibfield  {author} {\bibinfo {author} {\bibfnamefont {F.}~\bibnamefont
  {de~Felice}}\ and\ \bibinfo {author} {\bibfnamefont {Y.}~\bibnamefont {Yu}},\
  }\href@noop {} {\bibfield  {journal} {\bibinfo  {journal} {Class. Quant.
  Grav.}\ }\textbf {\bibinfo {volume} {18}},\ \bibinfo {pages} {1235} (\bibinfo
  {year} {2001})}\BibitemShut {NoStop}%
\bibitem [{\citenamefont {Hod}(2002)}]{hod2002}%
  \BibitemOpen
  \bibfield  {author} {\bibinfo {author} {\bibfnamefont {S.}~\bibnamefont
  {Hod}},\ }\href@noop {} {\bibfield  {journal} {\bibinfo  {journal} {Phys.
  Rev. D}\ }\textbf {\bibinfo {volume} {66}},\ \bibinfo {pages} {024016}
  (\bibinfo {year} {2002})}\BibitemShut {NoStop}%
\bibitem [{\citenamefont {Jacobson}\ and\ \citenamefont
  {Sotiriou}(2009)}]{jacobson2009}%
  \BibitemOpen
  \bibfield  {author} {\bibinfo {author} {\bibfnamefont {T.}~\bibnamefont
  {Jacobson}}\ and\ \bibinfo {author} {\bibfnamefont {T.}~\bibnamefont
  {Sotiriou}},\ }\href@noop {} {\bibfield  {journal} {\bibinfo  {journal}
  {Phys. Rev. Lett.}\ }\textbf {\bibinfo {volume} {103}},\ \bibinfo {pages}
  {141101} (\bibinfo {year} {2009})}\BibitemShut {NoStop}%
\bibitem [{\citenamefont {Chirco}\ \emph {et~al.}(2010)\citenamefont {Chirco},
  \citenamefont {Liberati},\ and\ \citenamefont {Sotiriou}}]{sotiriou2010}%
  \BibitemOpen
  \bibfield  {author} {\bibinfo {author} {\bibfnamefont {G.}~\bibnamefont
  {Chirco}}, \bibinfo {author} {\bibfnamefont {S.}~\bibnamefont {Liberati}}, \
  and\ \bibinfo {author} {\bibfnamefont {T.~P.}\ \bibnamefont {Sotiriou}},\
  }\href@noop {} {\bibfield  {journal} {\bibinfo  {journal} {Phys. Rev. D}\
  }\textbf {\bibinfo {volume} {82}},\ \bibinfo {pages} {104015} (\bibinfo
  {year} {2010})}\BibitemShut {NoStop}%
\bibitem [{\citenamefont {Saa}\ and\ \citenamefont
  {Santarelli}(2011)}]{saa2011}%
  \BibitemOpen
  \bibfield  {author} {\bibinfo {author} {\bibfnamefont {A.}~\bibnamefont
  {Saa}}\ and\ \bibinfo {author} {\bibfnamefont {R.}~\bibnamefont
  {Santarelli}},\ }\href@noop {} {\bibfield  {journal} {\bibinfo  {journal}
  {Phys. Rev. D}\ }\textbf {\bibinfo {volume} {84}},\ \bibinfo {pages} {027501}
  (\bibinfo {year} {2011})}\BibitemShut {NoStop}%
\bibitem [{\citenamefont {Gao}\ and\ \citenamefont {Zhang}(2013)}]{gao2013}%
  \BibitemOpen
  \bibfield  {author} {\bibinfo {author} {\bibfnamefont {S.}~\bibnamefont
  {Gao}}\ and\ \bibinfo {author} {\bibfnamefont {Y.}~\bibnamefont {Zhang}},\
  }\href@noop {} {\bibfield  {journal} {\bibinfo  {journal} {Phys. Rev. D}\
  }\textbf {\bibinfo {volume} {87}},\ \bibinfo {pages} {044028} (\bibinfo
  {year} {2013})}\BibitemShut {NoStop}%
\bibitem [{\citenamefont {Zimmerman}\ \emph {et~al.}(2013)\citenamefont
  {Zimmerman}, \citenamefont {Vega}, \citenamefont {Poisson},\ and\
  \citenamefont {Haas}}]{poisson2013}%
  \BibitemOpen
  \bibfield  {author} {\bibinfo {author} {\bibfnamefont {P.}~\bibnamefont
  {Zimmerman}}, \bibinfo {author} {\bibfnamefont {I.}~\bibnamefont {Vega}},
  \bibinfo {author} {\bibfnamefont {E.}~\bibnamefont {Poisson}}, \ and\
  \bibinfo {author} {\bibfnamefont {R.}~\bibnamefont {Haas}},\ }\href@noop {}
  {\bibfield  {journal} {\bibinfo  {journal} {Phys. Rev. D}\ }\textbf {\bibinfo
  {volume} {87}},\ \bibinfo {pages} {041501} (\bibinfo {year}
  {2013})}\BibitemShut {NoStop}%
\bibitem [{\citenamefont {Barausse}\ \emph {et~al.}(2010)\citenamefont
  {Barausse}, \citenamefont {Cardoso},\ and\ \citenamefont
  {Khanna}}]{barausse2010}%
  \BibitemOpen
  \bibfield  {author} {\bibinfo {author} {\bibfnamefont {E.}~\bibnamefont
  {Barausse}}, \bibinfo {author} {\bibfnamefont {V.}~\bibnamefont {Cardoso}}, \
  and\ \bibinfo {author} {\bibfnamefont {G.}~\bibnamefont {Khanna}},\
  }\href@noop {} {\bibfield  {journal} {\bibinfo  {journal} {Phys. Rev. Lett.}\
  }\textbf {\bibinfo {volume} {105}},\ \bibinfo {pages} {261102} (\bibinfo
  {year} {2010})}\BibitemShut {NoStop}%
\bibitem [{\citenamefont {Barausse}\ \emph {et~al.}(2011)\citenamefont
  {Barausse}, \citenamefont {Cardoso},\ and\ \citenamefont
  {Khanna}}]{barausse2011}%
  \BibitemOpen
  \bibfield  {author} {\bibinfo {author} {\bibfnamefont {E.}~\bibnamefont
  {Barausse}}, \bibinfo {author} {\bibfnamefont {V.}~\bibnamefont {Cardoso}}, \
  and\ \bibinfo {author} {\bibfnamefont {G.}~\bibnamefont {Khanna}},\
  }\href@noop {} {\bibfield  {journal} {\bibinfo  {journal} {Phys. Rev. D}\
  }\textbf {\bibinfo {volume} {84}},\ \bibinfo {pages} {104006} (\bibinfo
  {year} {2011})}\BibitemShut {NoStop}%
\bibitem [{\citenamefont {Colleoni}\ and\ \citenamefont
  {Barack}(2015)}]{barack2015A}%
  \BibitemOpen
  \bibfield  {author} {\bibinfo {author} {\bibfnamefont {M.}~\bibnamefont
  {Colleoni}}\ and\ \bibinfo {author} {\bibfnamefont {L.}~\bibnamefont
  {Barack}},\ }\href@noop {} {\bibfield  {journal} {\bibinfo  {journal} {Phys.
  Rev. D}\ }\textbf {\bibinfo {volume} {91}},\ \bibinfo {pages} {104024}
  (\bibinfo {year} {2015})}\BibitemShut {NoStop}%
\bibitem [{\citenamefont {Colleoni}\ \emph {et~al.}(2015)\citenamefont
  {Colleoni}, \citenamefont {Barack}, \citenamefont {Shah},\ and\ \citenamefont
  {van~de Meent}}]{barack2015B}%
  \BibitemOpen
  \bibfield  {author} {\bibinfo {author} {\bibfnamefont {M.}~\bibnamefont
  {Colleoni}}, \bibinfo {author} {\bibfnamefont {L.}~\bibnamefont {Barack}},
  \bibinfo {author} {\bibfnamefont {A.~G.}\ \bibnamefont {Shah}}, \ and\
  \bibinfo {author} {\bibfnamefont {M.}~\bibnamefont {van~de Meent}},\
  }\href@noop {} {\bibfield  {journal} {\bibinfo  {journal} {Phys. Rev. D}\
  }\textbf {\bibinfo {volume} {92}},\ \bibinfo {pages} {084044} (\bibinfo
  {year} {2015})}\BibitemShut {NoStop}%
\bibitem [{\citenamefont {Gao}\ and\ \citenamefont {Wald}(2001)}]{wald2001}%
  \BibitemOpen
  \bibfield  {author} {\bibinfo {author} {\bibfnamefont {S.}~\bibnamefont
  {Gao}}\ and\ \bibinfo {author} {\bibfnamefont {R.~M.}\ \bibnamefont {Wald}},\
  }\href@noop {} {\bibfield  {journal} {\bibinfo  {journal} {Phys. Rev. D}\
  }\textbf {\bibinfo {volume} {64}},\ \bibinfo {pages} {084020} (\bibinfo
  {year} {2001})}\BibitemShut {NoStop}%
\bibitem [{\citenamefont {T\'{o}th}(2012)}]{toth2012}%
  \BibitemOpen
  \bibfield  {author} {\bibinfo {author} {\bibfnamefont {G.~Z.}\ \bibnamefont
  {T\'{o}th}},\ }\href@noop {} {\bibfield  {journal} {\bibinfo  {journal} {Gen.
  Relativ. Gravit.}\ }\textbf {\bibinfo {volume} {44}},\ \bibinfo {pages}
  {2019} (\bibinfo {year} {2012})}\BibitemShut {NoStop}%
\bibitem [{\citenamefont {Nat\'{a}rio}\ \emph {et~al.}(2016)\citenamefont
  {Nat\'{a}rio}, \citenamefont {Queimada},\ and\ \citenamefont
  {Vicente}}]{natario2016}%
  \BibitemOpen
  \bibfield  {author} {\bibinfo {author} {\bibfnamefont {J.}~\bibnamefont
  {Nat\'{a}rio}}, \bibinfo {author} {\bibfnamefont {L.}~\bibnamefont
  {Queimada}}, \ and\ \bibinfo {author} {\bibfnamefont {R.}~\bibnamefont
  {Vicente}},\ }\href@noop {} {\bibfield  {journal} {\bibinfo  {journal}
  {Class. Quant. Grav.}\ }\textbf {\bibinfo {volume} {33}},\ \bibinfo {pages}
  {175002} (\bibinfo {year} {2016})}\BibitemShut {NoStop}%
\bibitem [{\citenamefont {Gralla}\ and\ \citenamefont
  {Wald}(2011)}]{gralla_wald_1}%
  \BibitemOpen
  \bibfield  {author} {\bibinfo {author} {\bibfnamefont {S.~E.}\ \bibnamefont
  {Gralla}}\ and\ \bibinfo {author} {\bibfnamefont {R.~M.}\ \bibnamefont
  {Wald}},\ }\href@noop {} {\bibfield  {journal} {\bibinfo  {journal} {Class.
  Quant. Grav.}\ }\textbf {\bibinfo {volume} {28}},\ \bibinfo {pages} {159501}
  (\bibinfo {year} {2011})}\BibitemShut {NoStop}%
\bibitem [{\citenamefont {Gralla}\ \emph {et~al.}(2009)\citenamefont {Gralla},
  \citenamefont {Harte},\ and\ \citenamefont {Wald}}]{gralla_wald_2}%
  \BibitemOpen
  \bibfield  {author} {\bibinfo {author} {\bibfnamefont {S.~E.}\ \bibnamefont
  {Gralla}}, \bibinfo {author} {\bibfnamefont {A.~I.}\ \bibnamefont {Harte}}, \
  and\ \bibinfo {author} {\bibfnamefont {R.~M.}\ \bibnamefont {Wald}},\
  }\href@noop {} {\bibfield  {journal} {\bibinfo  {journal} {Phys. Rev. D}\
  }\textbf {\bibinfo {volume} {80}},\ \bibinfo {pages} {024031} (\bibinfo
  {year} {2009})}\BibitemShut {NoStop}%
\bibitem [{\citenamefont {Hollands}\ and\ \citenamefont
  {Wald}(2013)}]{wald2012}%
  \BibitemOpen
  \bibfield  {author} {\bibinfo {author} {\bibfnamefont {S.}~\bibnamefont
  {Hollands}}\ and\ \bibinfo {author} {\bibfnamefont {R.~M.}\ \bibnamefont
  {Wald}},\ }\href@noop {} {\bibfield  {journal} {\bibinfo  {journal} {Commun.
  Math. Phys.}\ }\textbf {\bibinfo {volume} {321}},\ \bibinfo {pages} {629}
  (\bibinfo {year} {2013})}\BibitemShut {NoStop}%
\bibitem [{\citenamefont {Wald}(1984)}]{wald_book}%
  \BibitemOpen
  \bibfield  {author} {\bibinfo {author} {\bibfnamefont {R.~M.}\ \bibnamefont
  {Wald}},\ }\href@noop {} {\emph {\bibinfo {title} {General Relativity}}}\
  (\bibinfo  {publisher} {University of Chicago Press},\ \bibinfo {year}
  {1984})\BibitemShut {NoStop}%
\bibitem [{\citenamefont {Prabhu}(2017)}]{prabhu2017}%
  \BibitemOpen
  \bibfield  {author} {\bibinfo {author} {\bibfnamefont {K.}~\bibnamefont
  {Prabhu}},\ }\href@noop {} {\bibfield  {journal} {\bibinfo  {journal} {Class.
  Quant. Grav.}\ }\textbf {\bibinfo {volume} {34}},\ \bibinfo {pages} {035011}
  (\bibinfo {year} {2017})}\BibitemShut {NoStop}%
\bibitem [{\citenamefont {Iyer}\ and\ \citenamefont {Wald}(1994)}]{wald1994}%
  \BibitemOpen
  \bibfield  {author} {\bibinfo {author} {\bibfnamefont {V.}~\bibnamefont
  {Iyer}}\ and\ \bibinfo {author} {\bibfnamefont {R.~M.}\ \bibnamefont
  {Wald}},\ }\href@noop {} {\bibfield  {journal} {\bibinfo  {journal} {Phys.
  Rev. D}\ }\textbf {\bibinfo {volume} {50}},\ \bibinfo {pages} {846} (\bibinfo
  {year} {1994})}\BibitemShut {NoStop}%
\bibitem [{\citenamefont {Iyer}\ and\ \citenamefont {Wald}(1995)}]{wald1995}%
  \BibitemOpen
  \bibfield  {author} {\bibinfo {author} {\bibfnamefont {V.}~\bibnamefont
  {Iyer}}\ and\ \bibinfo {author} {\bibfnamefont {R.~M.}\ \bibnamefont
  {Wald}},\ }\href@noop {} {\bibfield  {journal} {\bibinfo  {journal} {Phys.
  Rev. D}\ }\textbf {\bibinfo {volume} {52}},\ \bibinfo {pages} {4430}
  (\bibinfo {year} {1995})}\BibitemShut {NoStop}%
\bibitem [{\citenamefont {Carter}(1973)}]{carter_book}%
  \BibitemOpen
  \bibfield  {author} {\bibinfo {author} {\bibfnamefont {B.}~\bibnamefont
  {Carter}},\ }in\ \href@noop {} {\emph {\bibinfo {booktitle} {Les Houches
  1972}}},\ \bibinfo {editor} {edited by\ \bibinfo {editor} {\bibfnamefont
  {C.}~\bibnamefont {Dewitt}}\ and\ \bibinfo {editor} {\bibfnamefont {B.~S.}\
  \bibnamefont {Dewitt}}}\ (\bibinfo  {publisher} {Gordon and Breach},\
  \bibinfo {address} {New York},\ \bibinfo {year} {1973})\BibitemShut {NoStop}%
\bibitem [{\citenamefont {Christodoulou}(1970)}]{christodoulou1970}%
  \BibitemOpen
  \bibfield  {author} {\bibinfo {author} {\bibfnamefont {D.}~\bibnamefont
  {Christodoulou}},\ }\href@noop {} {\bibfield  {journal} {\bibinfo  {journal}
  {Phys. Rev. Lett.}\ }\textbf {\bibinfo {volume} {25}},\ \bibinfo {pages}
  {1596} (\bibinfo {year} {1970})}\BibitemShut {NoStop}%
\bibitem [{\citenamefont {Aretakis}(2012)}]{aretakis}%
  \BibitemOpen
  \bibfield  {author} {\bibinfo {author} {\bibfnamefont {S.}~\bibnamefont
  {Aretakis}},\ }\href@noop {} {} (\bibinfo {year} {2012}),\ \Eprint
  {http://arxiv.org/abs/1206.6598} {arXiv:1206.6598 [gr-qc]} \BibitemShut
  {NoStop}%
\bibitem [{\citenamefont {L\'{e}aut\'{e}}\ and\ \citenamefont
  {Linet}(1982)}]{linet1982}%
  \BibitemOpen
  \bibfield  {author} {\bibinfo {author} {\bibfnamefont {B.}~\bibnamefont
  {L\'{e}aut\'{e}}}\ and\ \bibinfo {author} {\bibfnamefont {B.}~\bibnamefont
  {Linet}},\ }\href@noop {} {\bibfield  {journal} {\bibinfo  {journal} {J.
  Phys. A.}\ }\textbf {\bibinfo {volume} {15}},\ \bibinfo {pages} {1821}
  (\bibinfo {year} {1982})}\BibitemShut {NoStop}%
\bibitem [{\citenamefont {Unruh}\ and\ \citenamefont {Wald}(1982)}]{wald1982}%
  \BibitemOpen
  \bibfield  {author} {\bibinfo {author} {\bibfnamefont {W.~G.}\ \bibnamefont
  {Unruh}}\ and\ \bibinfo {author} {\bibfnamefont {R.~M.}\ \bibnamefont
  {Wald}},\ }\href@noop {} {\bibfield  {journal} {\bibinfo  {journal} {Phys.
  Rev. D}\ }\textbf {\bibinfo {volume} {25}},\ \bibinfo {pages} {942} (\bibinfo
  {year} {1982})}\BibitemShut {NoStop}%
\bibitem [{\citenamefont {Boulware}(1973)}]{boulware1973}%
  \BibitemOpen
  \bibfield  {author} {\bibinfo {author} {\bibfnamefont {D.~G.}\ \bibnamefont
  {Boulware}},\ }\href@noop {} {\bibfield  {journal} {\bibinfo  {journal}
  {Phys. Rev. D}\ }\textbf {\bibinfo {volume} {8}},\ \bibinfo {pages} {2363}
  (\bibinfo {year} {1973})}\BibitemShut {NoStop}%
\end{thebibliography}
\end{document}